\documentclass[fleqn,11pt]{wlscirep}
\usepackage[utf8]{inputenc}
\usepackage[T1]{fontenc}
\usepackage{graphicx}
\usepackage{epstopdf}
\usepackage{dcolumn}
\usepackage{bm}
\usepackage{braket}
\usepackage{color}
\usepackage{ragged2e}
\usepackage{amsmath, bm}
\usepackage{amsmath}
\usepackage{amssymb}%
\usepackage{url} %
\usepackage{tcolorbox}
\usepackage{hyperref}
\usepackage{bookmark}
\usepackage{physics}

\newcommand{\Blue}[1]{\textcolor{black}{#1}}
\newcommand{\Red}[1]{\textcolor{black}{#1}}

\title{Engineering spectro-temporal light states with physics-embedded deep learning }
\author[1,2,*]{Shilong Liu }
\author[1]{St\'{e}phane Virally}
\author[1]{Gabriel Demontigny}
\author[1]{Patrick Cusson}
\author[1,+]{Denis V. Seletskiy}

\affil[1]{femtoQ Lab, Department of Engineering Physics, Polytechnique Montr\'{e}al, Montr\'{e}al, Qu\'{e}bec H3T 1J4, Canada}
\affil[2]{Tempo Optics Inc., Montr\'{e}al, Qu\'{e}bec H3T 1W9, Canada}

\affil[*]{corresponding dr.shilongliu@gmail.com}

\affil[+]{corresponding  denis.seletskiy@polymtl.ca}

\begin{abstract}
Frequency synthesis and spectro-temporal control of optical wave packets are central to ultrafast science, with supercontinuum (SC) generation standing as one remarkable example. Through passive manipulation, femtosecond (fs) pulses from nJ-level lasers can be transformed into octave-spanning spectra, supporting few-cycle pulse outputs when coupled with external pulse compressors. While strategies such as machine learning have been applied to control the SC's central wavelength and bandwidth, their success has been limited by the nonlinearities and strong sensitivity to measurement noise.
Here, we propose and demonstrate how a \Red{physics-embedded} convolutional neural network (P-CNN) that embeds spectro-temporal correlations can circumvent such challenges, resulting in faster convergence and reduced noise sensitivity. This innovative approach enables on-demand control over spectro-temporal features of SC, achieving few-cycle pulse shaping without external compressors. This approach heralds a new era of arbitrary spectro-temporal light state engineering, with implications for ultrafast photonics,  photonic neuromorphic computation, and AI-driven optical systems.
 \end{abstract}

\begin{document}
\flushbottom
\maketitle
\thispagestyle{empty}
\noindent

Supercontinuum (SC) is a nonlinear optical phenomenon by which a spectrum of a  femtosecond (fs) input pulse becomes broadened by one or two orders of magnitude, often extending beyond an optical octave. It arises from the intricate interplay of third-order and higher-order nonlinear optical processes when the fs pulse propagates through a dispersive medium. Femtosecond laser oscillators delivering pulses with energies in the nanojoule (nJ) range can drive SC generation in highly nonlinear media such as photonic-crystal fibers or bulk glass \cite{Ranka2000OL,agrawal2000nonlinear,alfano2006supercontinuum,dudley2006supercontinuum,Sell2009OE}. The resulting SC significantly broadens the incident pulse spectrum, creating a continuous, octave-spanning spectral band. With high brightness and exceptional spectral coverage, SC  has become an indispensable resource for applications such as spectroscopy \cite{borondics2018supercontinuum}, advanced microscopy \cite{kaminski2008supercontinuum,gundougdu2023self}, and optical coherence tomography \cite{tu2013coherent,ji2021millimeter}. In ultrafast photonics, SC generation and precise control of their spectro-temporal features serves as a cornerstone for breakthroughs in ultrashort laser sources, including self-referenced \cite{Jones2000Sci,Sinclair2015comb} frequency combs \cite{Fehrenbacher2015Optica}, few- and single-cycle pulses \cite{Krauss2010NatPhot,shumakova2016multi,steinleitner2022single}, and high-harmonic generation as well as attosecond pulse \cite{Lewenstein1994PRA,Corkum2007NatPhys,Krausz2024RMP,peng2019attosecond,li2020attosecond}.

Although the precise manipulation of SC spectra is highly desirable, it poses significant experimental challenges. The essential difficulty arises from the extreme sensitivity of SC generation to complex dispersion and input pulse parameters, which drive a cascade of nonlinear interactions, including self- and cross-phase modulation (SPM and XPM), four-wave mixing (FWM), and Raman scattering. The task of directing these processes to produce predictable SC outputs is formidable, which requires detailed characterization of the parameters of the materials, such as dispersion, third-order nonlinearity \cite{alfano2006supercontinuum, Sell2009OE, bres2023supercontinuum}, and input temporal profile \cite{lozovoy2004multiphoton,xu2006quantitative,monmayrant2010newcomer}. Achieving such control could enable the routine generation of octave-spanning, thereby democratizing access to few-cycle pulse shaping \cite{Krauss2010NatPhot, steinleitner2022single}.

Various strategies have been developed to control over SC generation. One approach involves selecting materials or waveguides with favorable dispersion properties \cite{Sell2009OE,Carlson2017OL} or tailoring the non-linearity of the medium, for example, by adjusting the pressure inside gas-filled hollow core fibers \cite{beetar2020multioctave, elu2021seven, piccoli2021intense}. Other methods focus on modifying the input pulse specifications, such as pulse duration \cite{genty2007fiber}, energy \cite{genier2021ultra}, and pulse number \cite{wetzel2018customizing}, to adjust SC spectral properties \cite{dudley2006supercontinuum, sylvestre2021recent}. Incorporating optimization algorithms, particularly genetic algorithms (GA) in an iterative process, has enabled dynamic control of SC features \cite{tzang2018adaptive, wetzel2018customizing, hary2023tailored, lapre2023genetic}, in which the searching time typically takes several minutes.

Recent theoretical studies employing machine learning have shown promising capabilities in predicting the spectral or  temporal characteristics of SC \cite{martins2022design, hoang2022optimizing, shih2023maximizing}. The main feature of machine learning algorithms, such as feed-forward neural networks (FNNs), is the ability to learn the pulse evolution in the nonlinear complex systems, dramatically reducing the time toward optimized target functions from minutes to mere seconds   \cite{farfan2018femtosecond, boscolo2020artificial, salmela2022feed}. By leveraging machine learning, experimental investigations have analyzed specific SC properties, particularly identifying extreme events \cite{narhi2018machine, salmela2020machine}. These studies typically rely on data-driven models \cite{genty2021machine}, using phase and intensity inputs to train a neural network.

Despite these advances, SC generation presents a strongly underdetermined problem due to the dramatic spectral broadening relative to the input pulse. The spectro-temporal correlations of SC are highly sensitive to the input wave's spectral phase and amplitude, which evolves intricately during nonlinear propagation.
When small correlations become comparable to the experimental noise (or numerical instabilities) in the training datasets, the search for the optimized solution becomes compromised. Consequently, large and accurate training datasets are often required for adequate network convergence \cite{boscolo2020artificial,genty2021machine}.

In this work, we introduce  a \Red{physics-embedded} convolutional neural network (P-CNN) to address these challenges and perform spectro-temporal pulse shaping in nonlinear optics. We use the term "\Red{physics-embedded}" to underscore the fact that we embed the physical transform, i.e., spectro-temporal correlation function, specifically the Wigner function, into the front of network instead of relying solely on data-driven models. The Wigner function represents wavepackets in the spectral and temporal (chronocyclic) domain, enabling the CNNs to filter out uncorrelated noise efficiently, thus reducing sensitivity to phase variations such as carrier-envelope offset phase (CEP) \cite{zuo2022deep}. \Red{Simulation results indicate that the P-CNN approach achieves a convergence approximately three times better than the purely data-driven FNN. Furthermore, additional simulations demonstrate that even a simple FNN —when embedded with Wigner-function-based input (referred to as P-FNN)— can also achieve significantly improved performance. This underscores the importance of physics-embedded input representations, regardless of model complexity.}

Experimentally, we demonstrate a high nonlinear optical platform with implementation of two P-CNNs. 
Enabled by one P-CNN, 
the system shows continuous tuning of SC's spectral and temporal features, including central frequency, bandwidth, and pulse duration. Notably, we achieve 12 fs pulses (sub-three optical cycle; a 70-fold nonlinear compression of pulse duration) directly out of a SC fiber, thus eliminating the need for an external pulse compressor \cite{dudley2006supercontinuum, piccoli2021intense}. Using the second P-CNN, the system can directly output high-order soliton waveform over arbitrary spectral-temporal profile. These results underscore the physics-embedded network advancements in manipulation of SC, paving the way for precise and efficient control over  spectro-temporal laser state in highly nonlinear system.

\newpage

\section*{P-CNN framework via the nonlinear optical platform}

\begin{figure}
  \centering
    \includegraphics[width=17cm]{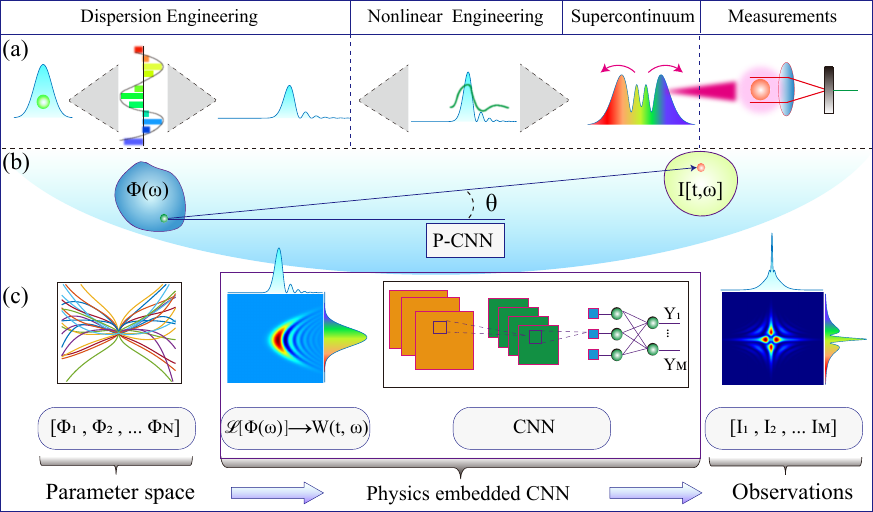}
       \caption{Framework for spectro-temporal engineering using  \Red{physics-embedded} convolutional neural network (P-CNN). (a) Schematic setup including dispersion and nonlinear engineering, supercontinuum generation, and measurement stages. (b) Mapping between input parameter spaces and observations via a nonlinear angle \(\theta\). (c) \Red{The architecture of the physics neural networks, incorporating parameters space, physics embedded CNN, and observations}
}\label{F1}
\end{figure}

Figure \ref{F1}(a) provides an overview of our nonlinear optical platform for implementation of P-CNN in the context of the SC generation. The setup consists of four primary sections. The first section, responsible for dispersion engineering, includes a mode-locked fiber laser that generates femtosecond soliton pulses and a 4$f$ pulse shaper, which enables precise complex dispersion control. The output pulse is fed into the second nonlinear engineering section, which comprises an Erbium-doped fiber amplifier (EDFA) to address pulse energy enhancement and self-phase modulation(SPM). The third section is a highly nonlinear fiber (HNF), which is used for supercontinuum (SC) generation. The spectrum (or temporal features) of SC is measured in the last section, using a pulse characterization setup. This platform allows for engineering of an octave-spanning spectral broadening of SC, unlocking advanced manipulation of few-cycle pulse shaping. For further details about the setup and simulations of SC generation, refer to \hyperref[sec:methods1]{\textbf{Methods-1: Setup and Simulations for Supercontinuum Generation}}.

The dynamics of a laser pulse with a temporal envelope $A(z,t)$ propagating through a HNF can be modeled by a Generalized Nonlinear Schrödinger Equation (GNLSE) \cite{dudley2006supercontinuum}:

\begin{equation}\label{E-1}
\qty[\pdv{z}-i\sum_{k\geq2}\frac{i^k\beta _k}{k!}\pdv[k]{t}+\frac\alpha2]A(z,t)=\gamma\qty(i-\tau_{\mathrm{shock}}\pdv{t})\qty[i\Gamma_R(z,t)+A(z,t)\int_{-\infty}^{+\infty}dt'\;R(t')\abs{A(z,t-t')}^2].
\end{equation}

Here, $z$ is the propagation coordinate and $t$ represents the retarded time $t=T-\beta_1z$ evolving with the group velocity (first-order dispersion) $\beta_1$ of the envelope. The left-hand side of the equation models linear propagation effects. The right-hand side accounts for nonlinear effects, in which the Kerr coefficient $\gamma$ governs the nonlinear refractive index, and $R(t)$ is the Raman response function. The term $\Gamma_R(z,t)$ captures the influence of spontaneous Raman noise. Additional nonlinear phenomena, such as self-steepening and optical shock formation, are characterized by a timescale $\tau_{shock} = \tau_0 = 1/\omega_0$.

Typically, the dynamics of SC generation is resolved through numerical simulations based on Eq.~\ref{E-1}. However, such simulations are computationally intensive and highly sensitive to the aforementioned parameters and the shape of the input pulse \(A_\text{in}(t)\), which are often derived experimentally \cite{Sell2009OE}. Each initial condition defines a unique SC output through a nonlinear mapping process. Since the input temporal field is generally defined in the spectral domain via the Fourier transform \(\mathcal{F}\), it can be expressed as \({A}_{\text{in}}(t) = \mathcal{F}^{-1}\left[\tilde A_{\text{in}}(\omega)\exp\left(i\Phi(\omega)\right)\right]\), where \(\tilde A_{\text{in}}(\omega)\) and \(\Phi(\omega)\) denote the spectral amplitude and phase, respectively. 
Thus, direct observation of the spectral or temporal intensity \(\text{I[t},\omega\text{]}\) of the SC provides experimental access to the targeted parameter space of \(\Phi(\omega)\). 

A robust program for engineering SC is to find a nonlinear mapping of an angle \(\theta\) between the parameter space and the observed output \(\text{I[t},\omega\text{]}\), as sketched in Fig.~\ref{F1}(b). Finding the accurate angle \(\theta\) can be quite challenging both in simulations of Eq.~\ref{E-1} and in experiments, due to the interplay of nonlinear processes and corresponding sensitivity to experimental (or computational) noise. Machine learning algorithms present a promising solution by simplifying this complex nonlinear system into a manageable network representation. As shown in Fig.~\ref{F1}(c), a \Red{physics-embedded} convolutional neural network (P-CNN) is introduced to efficiently learn the nonlinear mapping, expressed as:

\begin{equation}\label{E-2}
      \text{I[t},\omega\text{]}=\text{P-CNN}_{\rm{\theta}}[\Phi(\omega)]=\rm{CNN}[\mathcal{T}(\Phi(\omega))]=\rm{CNN}[W(t,\omega)],
\end{equation}
where the P-CNN is based on the CNN embedded with the physical transform \(\mathcal{T}(\Phi(\omega))\). The physical transform, here the Wigner function, is designed to represent the pulse in the chronocyclic domain\cite{walmsley2009characterization,najafabadi2024intensity}, containing the structure of spectro-temporal correlations:
\begin{equation}\label{E-3}
   W(t,\omega) = \int_{-\infty}^{+\infty}dt'\; A\qty(t + \frac{t'}{2}) A^*\qty(t - \frac{t'}{2})\,e^{i\omega t'}.
\end{equation}
Here $A(t)$ is associated with the spectral amplitude $\tilde A(\omega)$ and phase $\Phi(\omega)$.
Experimentally, such pulse profile can be reconstructed by using time-dependent spectrograms, such as Second-Harmonic-Generation Frequency-Resolved Optical Gating (SHG-FROG). The success of the mapping (Eq. \ref{E-2}) hinges on CNN's ability to robustly identify the nonlinear transformation $\theta$, and thus find the optimal inverse solutions for SC engineering, with further details provided in \hyperref[sec:methods2]{\textbf{Methods-2: The Structure and Performance of Deep Learning Neural Networks}}.

\section*{Benchmarking P-CNN performance}
\begin{figure}
  \centering
  \includegraphics[width=17cm]{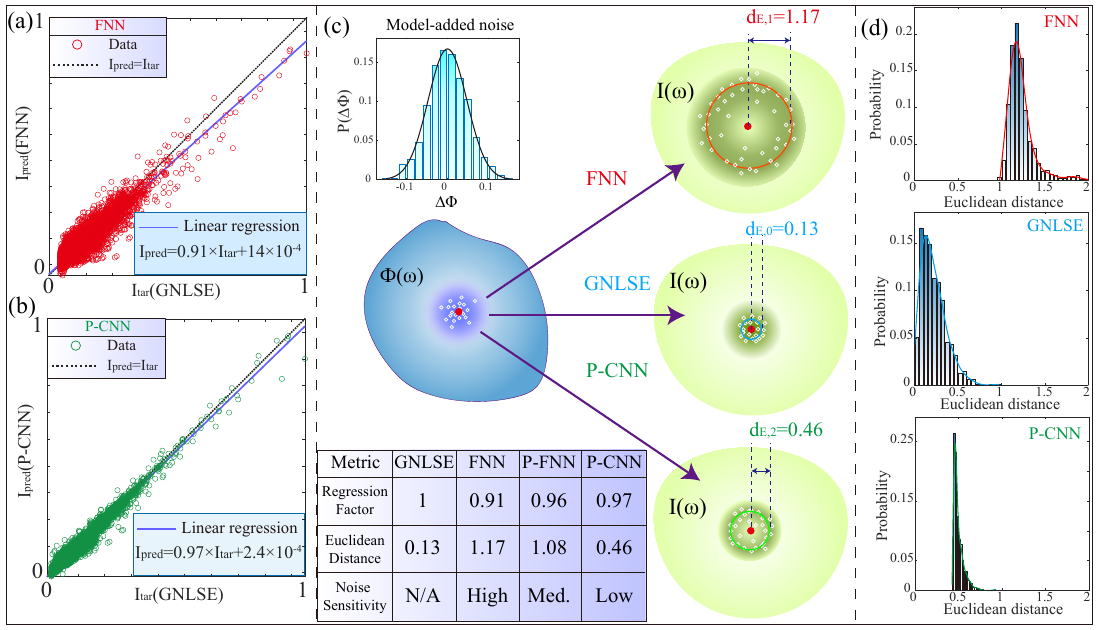}
  \caption{ Performance of \Red{physics embedded networks}. (a) and (b) Linear regression analysis for the predicted \(\mathrm{I_{pred}}\) from the FNN and P-CNN against the target \(\mathrm{I_{tar}}\) obtained from GNLSE simulations.
(c) Schematic representation of the phase \(\Phi(\omega)\) with added noise \(\Delta \Phi(\omega)\) (top-left panel) and its impact on the SC outputs (\(I(\omega)\)) across three approaches: FNN (top), GNLSE (middle), and P-CNN (bottom). The Euclidean distance (\(d\)) with maximum probability quantifies the deviation of the noisy SC output from the noiseless target spectrum, with values of \(\bar{d}_{E,0} = 0.13\), \(\bar{d}_{E,1} = 1.17\), \(\bar{d}_{E,0} = 1.08\) , and \(\bar{d}_{E,2} = 0.46\), respectively, for GNLSE, FNN, P-FNN and P-CNN. (d) Statistical distributions of the Euclidean distance for 1000 noisy phase inputs.
}\label{F1-2}
\end{figure}

To test the relative performance of the P-CNN, we first benchmark it against the FNN approach and in presence of simulated noise, while also comparing both to the exact solutions from the GNLSE (Eq.~\ref{E-1}). The training process of P-CNN begins with Latin Hypercube Sampling (LHS) \cite{mckay2000comparison} to scan the parameter space $\{X_n\}$ and generate diverse random phase profiles $\Phi(\omega)$. To enhance the diversity of the parameter space, we incorporate both regular and fractional dispersion controlled via using one 4$f$ pulse shaper (see fractional dispersion details in Refs.~\citenum{malomed2021optical,liu2023experimental}, also \textbf{Supplementary- S1} for additional details). For the observation targets, we set the spectral intensity $\text{I}(\omega)$ simulated from GNLSE as the objective function.

After training the network, we conducted a linear regression analysis of the SC outputs by comparing the target spectral intensity $\mathrm{I_{tar}{(\omega)}}$ from the GNLSE with the predicted intensity $\mathrm{I_{pred}(\omega)}$ from the trained networks. It should be noted that solving GNLSE numerically for every case is a computationally intensive task, not suitable for real-time applications. However, in does serve as an ideal benchmarking tool for our purpose.

\Red{We first trained a FNN with three hidden layers containing 200, 200, and 100 neurons, respectively. This configuration is optimized and commonly employed for SC predictions, as seen in prior studies \cite{boscolo2020artificial, salmela2022feed}. Figure~\ref{F1-2}(a) presents the correlation for the testing results of FNN, yielding a regression factor of $R = 0.91$ for the fitting line. The regression factor could be up to 0.96 when we embed the Wigner function data to the front of FNN, which refers to  P-FNN. For comparison, using the same training data, the regression results for the trained P-CNN, shown in Fig.~\ref{F1-2}(b), yield a higher correlation factor of $R \approx 0.97$.  Interpreting $(1-R)$ as the average prediction loss, the P-FNN and P-CNN demonstrates significantly superior performance, with around threefold reduction in loss compared to the FNN. }This is also reflected in the significantly lower variation of $\text{I}_{\text{pred}}(\text{P-CNN})$ as compared to $\text{I}_{\text{pred}}(\text{FNN})$ in Fig. \ref{F1-2}(a) and (b).  \Red{The non-uniformity observed in the normalized output distribution (Figs. 2(a-b)) stems from the underlying simulation data, where low-intensity values occur more frequently than high-intensity peaks. This is a natural consequence of the broadband and structured nature of the supercontinuum field. A histogram analysis of the original data confirms this skew.}

To further test the stability of these networks, we evaluate them under noisy conditions by selecting a specific phase profile $\Phi(\omega)$ and adding $\Delta \Phi(\omega)$ noise to it (\textbf{see Supplementary- S2 for more details}) . Here, we test against a normally distributed noise, following Gaussian statistics (top left panel of Fig. \ref{F1-2}(c)). \Blue{The selected Gaussian phase noise as a representative and widely used model for smooth phase fluctuations in both simulation and experimental systems, e.g. thermal noise or quantum-limited noise. It serves as a baseline for testing generalization under continuous random distortions }

The added phase noise modifies the initial pulse field and, consequently, alters the spectrum of the SC, revealing the aforementioned high sensitivity to the details of the spectral phase. To quantify the resulting variations in the SC spectrum, both with and without noise, we utilize the Euclidean distance $d_E$:
\begin{equation}\label{E-E}
  d_{E}=\sqrt{\sum_{\omega}\left(\mathrm{I_{predN}(\omega)-I_{tarG}(\omega)}\right)^2},
\end{equation}
\Red{where $\mathrm{I_{predN}(\omega)}$ represents the predictions from the GNLSE, FNN, P-FNN, or P-CNN model under noisy excitation}. $\mathrm{I_{tarG}(\omega)}$ is the target spectrum, simulated using the GNLSE without noise, indicated by the red dot in the three $\mathrm{I(\omega)}$ panels of Fig. \ref{F1-2}(c).

Next, we run 1000 simulations and evaluate the Euclidean distance for each of the three models, corresponding histograms of $d_E$  are shown in Fig. \ref{F1-2}(d). The GNLSE-based simulation serves as a reference, achieving a smaller maximum likelihood Euclidean distance of $\bar{d}_{E,0} = 0.13$ (Fig. \ref{F1-2}(c), middle).
The FNN's larger susceptibility to noise is reflected in a nearly 9-fold increase in the Euclidean distance ($\bar{d}_{E,1} = 1.17$, top row of Fig. \ref{F1-2}(c)). In contrast, $\bar{d}_{E,2} = 0.46$ for P-CNN is a 2.5-fold increase in comparison to the GNLSE. \Red{And, P-FNN's statistics produce the Euclidean distance $\sim$ 1.08, which also shows better noise performance than FNN.}
Also, the narrower width of the $d_E$ distribution for P-CNN in comparison to FNN provides strong indication of its more robustness to noise variation or low sensitivity to noise. These results are also summarized as a table inserted on the bottom left panel of Fig. \ref{F1-2}(c).  Unlike the traditional FNN, the \Red{physics embedded networks leverage} physical insight of spectro-temporal correlations to substantially reduce prediction errors while maintaining computational efficiency that surpasses direct GNLSE simulations. \textbf{(More details about networks, noise analysis, and P-FNN are shown in Supplementary- S2)}.

\section*{Experimental implementations}
To implement these methods experimentally, we constructed two distinct P-CNN networks: "Network-Soliton" (Net-S) and "Network-Dispersive" (Net-D), as illustrated in Fig.~\ref{F2}(a). These networks are designed to handle different pulse energy regimes.
For pulse energies $\sim $ 2 nJ, a cascade nonlinear process generates SC characterized by an ultra-broadband dispersive wave, blueshifted into the normal dispersion regime. These results are reported below under "Dispersive Wave". For lower pulse energies, the EDFA facilitates nonlinear spectral broadening through SPM. This regime operates in the anomalous dispersion domain, leading to spectral sidelobe generation and "soliton" formation.

\section*{Spectral Domain Engineering: Dispersive Wave}
\begin{figure}
  \centering
  \includegraphics[width=17cm]{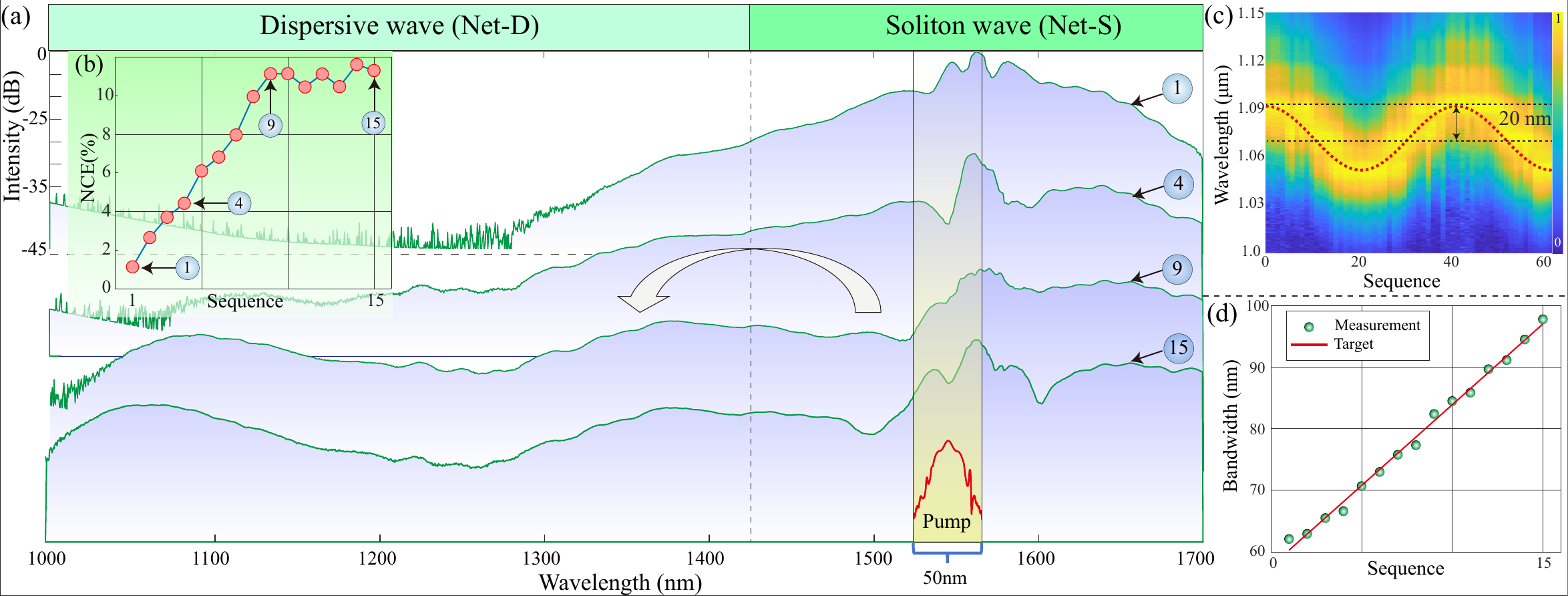}
  \caption{Spectral control of the supercontinuum. (a) Evolution of the SC spectra as a function of the input phase, controlled by the 4$f$ shaper. (b) Calculated nonlinear conversion efficiency defined in Eq. \ref{E_R} quantifies energy transfer from the soliton (Net-S) to dispersive wave (Net-D) spectral regions. The P-CNN enables engineering of the dispersive wave parameters: (c) The center wavelength oscillation with a target 3.7\% modulation around 1070 nm and (d) linear tunability in bandwidth for DW.}
  \label{F2}
\end{figure}

As a demonstration, we show the ability of P-CNN in dispersive wave (DW) engineering, including spectral manipulations and nonlinear conversion. 
To further enhance performance, a real-time gradient descent search algorithm was implemented within the real physics system, utilizing the optimized solutions (see \hyperref[sec:sup2]{\textbf{Supplementary - S3}} for details).

We first examine the controlled energy transfer from the soliton region (Net-S) to the dispersive wave region (Net-D). The Net-D region spans 1000–1430 nm, while the Net-S region covers 1430–1700 nm, constrained by the optical spectrum analyzer’s (OSA) range. Figure~\ref{F2}(a) shows the measured dependence of the SC output on the input spectral phase in log-linear scaling. Starting with modest broadening dominated by SPM (curve 1), P-CNN-assisted phase optimization progressively redistributes energy from soliton to dispersive wave (curves 4, 9, and 15). This process is quantified using the nonlinear conversion efficiency (NCE):

\begin{equation}\label{E_R}
  \eta_{NCE} = \frac{S_d}{S_d + S_s} \in [0, 1],
\end{equation}
where $S_{d,s} = \int_{d,s} I(\lambda) d\lambda$ represents the integrated spectral energy over the respective spectral regions. An $\eta_{NCE}$ of 1 indicates complete energy transfer to the dispersion wave, while 0 signifies no transfer. Figure~\ref{F2}(b) depicts the measured NCE, showing linear growth followed by oscillations, with a maximum conversion of 12\%.  In the high-NCE, we examine the DW centered around $\sim$1070 nm, which forms due to nonlinear interactions in the normal dispersion regime.

Next, we demonstrate the P-CNN's ability to manipulate the central wavelength and bandwidth. Regarding the central wavelength, $\lambda_{DS}$, the target is set by:
\begin{equation}\label{E_creep}
  \lambda_{DS} = \lambda_0 + \Delta\lambda\cos(\Theta),
\end{equation}
where $\lambda_0$ is the baseline wavelength, $\Delta\lambda$ is the modulation amplitude, and $\Theta$ is the angular coordinate. Figure~\ref{F2}(c) shows excellent agreement between the target modulation (solid blue line, $\Delta\lambda$ = 20 nm, $\lambda_0$ = 1070 nm) and measured data, achieving a 3.7\% modulation of the DW center wavelength.

We also demonstrate bandwidth manipulation with the DW center wavelength clamped at 1070 nm. The measured bandwidth closely follows the target range of 60-100 nm, as shown in Fig.~\ref{F2}(d). These combined results demonstrate the feasibility of a full engineering of the spectral features of the DW, while also tracing out a path toward an elegant control over the temporal domain, such as the few-cycle temporal wavepacket.

\section*{Temporal Domain Engineering: Few-cycle Pulse}

The DW is a useful resource to generate short-wavelength laser pulses \cite{agrawal2000nonlinear,kottig2017mid}, and the flexibility demonstrated in the previous section highlights its potential for precise control of center wavelength and bandwidth. The flattened and broad spectral feature observed in Fig.~\ref{F2}(a) further indicates the possibility of preparing ultrashort laser pulses. To isolate the DW from the long-wavelength spectral components of the Net-S band, we first employ a short-pass filter with a 1500 nm cutoff. Conventionally, this isolated DW is passed through dispersion-compensating elements, such as prism compressors or chirped mirrors, to produce near-transform-limited pulses. Instead, we leverage the P-CNN to simultaneously optimize the broadband DW generation and its compression, eliminating the need for additional compensators.

\begin{figure}
  \centering
  \includegraphics[width=17cm]{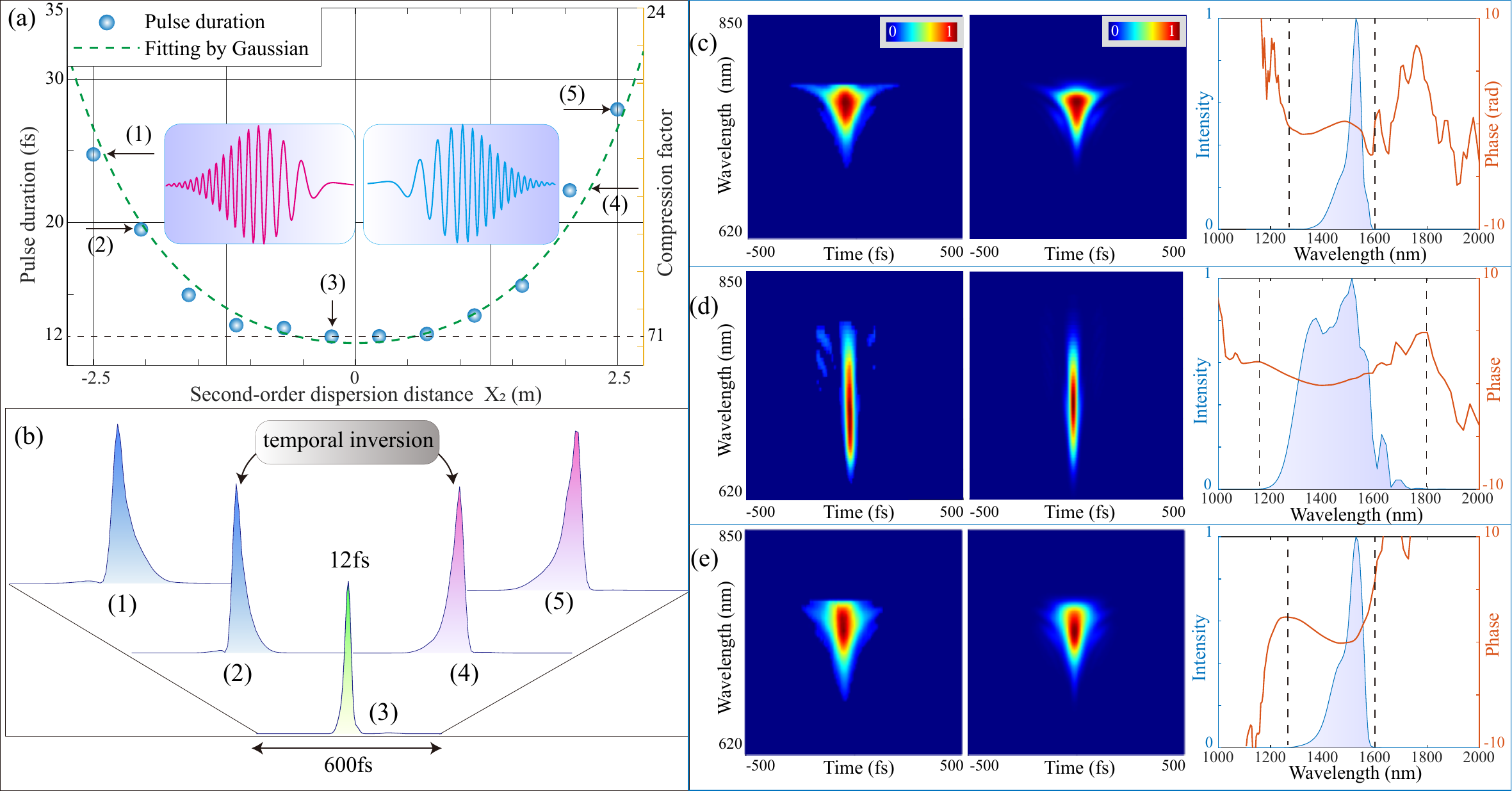}
  \caption{Temporal pulse duration engineering.  (a) Ordinates the measured relationship between the pulse duration (left ordinate axis) and second-order dispersion distance ($X_2$), with the right axis displaying the nonlinear compression factor.
(b)  The reconstructed temporal intensity at five positions marked in (a).
(c)-(e) The measured and reconstructed FROG traces for positions (1), (3), and (5) as indicated in panel (a), where the reconstructed spectral amplitude and phase are provided in the corresponding right-side panels.
  }\label{F3}
\end{figure}

SHG-FROG measurements are performed to evaluate the resulting pulse duration and profiles. The P-CNN is tasked to minimize the pulse duration, which is directly recognized through its Wigner-function-based training. For the optimized phase, the measured pulse duration is approximately 12 fs, corresponding to a compression factor of over 70, referenced as the soliton pulse duration of 850 fs. To explore temporal compression dynamics, the second-order dispersion length ($X_2$) is systematically adjusted from $-2.5$ to $2.5$ m. The corresponding pulse durations, shown in Fig.~\ref{F3}(a), also show the symmetric distribution, with a Gaussian scaling behavior (solid green line). On the right axis of Fig.~\ref{F3}(a), we add the calculated compression factor, illustrating the compression value from 24 to 71.

The evolution of temporal profiles retrieved from FROG measurements is shown in Fig.~\ref{F3}(b). The data reveal preserved inversion symmetry in the wavepacket envelope ($X_2 \leftrightarrow -X_2$) and opposite orientations of pulse tails for the same $|X_2|$, reminiscent of a "temporal inversion" effect \cite{moussa2023observation}. This symmetry, supported by GNLSE simulations (\hyperref[sec:sup4]{\textbf{Supplementary-S4}}), underscores the ability to shape temporal profiles through phase optimization.

Figures~\ref{F3}(c)-(e) present the measured and reconstructed FROG traces at positions (1), (3), and (5) from Fig.~\ref{F3}(a) and (b), alongside the corresponding reconstructed spectral amplitude and phase profiles. Notably, the spectral phase for positions (1) and (5) is opposite, resulting in symmetric structures in the temporal profiles. For positive and negative dispersion, symmetric self-steepening of the pulse is also observed, where a steeper trailing edge implies the potential to prepare arbitrary asymmetric few-cycle laser pulses \cite{anderson1983nonlinear}.

Further analysis reveals the influence of higher-order dispersion, such as fourth-order dispersion, on pulse duration symmetry. For the shortest pulse duration of $\sim$12 fs (Fig.~\ref{F3}(d)), the reconstructed spectral phase exhibits a residual second-order contribution near the central wavelength, corresponding to a group delay dispersion (GDD) of $\sim 100~\mathrm{fs^2}$- primarily arising from additional GDD in the setups of FROG that is estimated at around 120 $\mathrm{fs^2}$. By flattening the spectral phase, a transform-limited pulse duration of 8.5 fs is achieved. This small systematic offset can be mitigated by optimizing the dispersion in the FROG setup.  \Blue{Since the P-CNN was trained on target fields excluding measurement-system dispersion, it does not compensate for this post-shaping GDD. Future implementations may include a calibration routine or augmented training to address measurement-induced dispersion directly.}

Additional details on the FROG setup are provided in \hyperref[sec:sup5]{\textbf{Supplementary- S5}}.

\section*{Spectro-Temporal Engineering: High-Order Solitons}

In this section, we demonstrate the capability of the P-CNN to perform nonlinear pulse shaping in both the spectral and temporal domains. Here, the focus shifts back to the soliton Net-S region, previously defined. Since the pump resides within this spectral region, more complex patterns emerge, making on-demand shaping particularly challenging. However, the proposed P-CNN effectively balances the interplay between dispersion and nonlinear effects through optimized phase hologram in the pulse shaper.

\begin{figure}
  \centering
  \includegraphics[width=17cm]{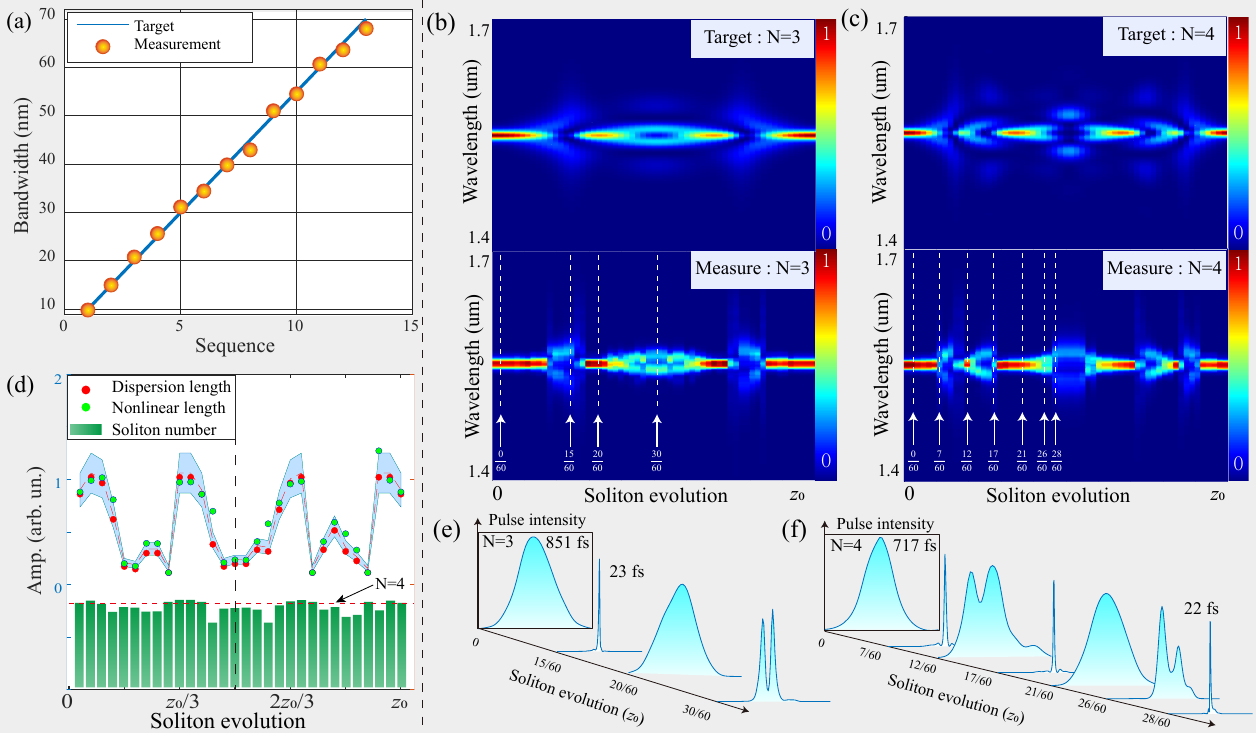}
  \caption{Spectro-temporal correlations for high-order soliton formation. (a) Bandwidth (3dB) of target functions (P-CNN) and measurements, ranging from 10 to 70 nm. (b)-(c) Spectra of high-order solitons for soliton numbers $N$=3 and $N$=4, with the top panels showing simulated target states and the bottom panels presenting experimental measurements. (d) Dispersion length, nonlinear length, and soliton number for a soliton with $N$=4. (e)-(f) Temporal intensity profiles reconstructed via FROG for various soliton sequences.}
  \label{F4}
\end{figure}

First, we demonstrate the optimization of soliton pulses with variable bandwidth for the soliton pulse. To achieve this, the P-CNN is trained in the Net-S region. The target bandwidth is linearly set from 10 to 70 nm, depicted by the blue line in Fig.~\ref{F4}(a), while the optimized measurement outcomes are represented  as orange dots. The results reveal prominent agreement between the measurements and the set targets, highlighting the system’s ability for precise and continuous tuning of the output's spectral bandwidth.

Next, we explore the feasibility to directly generate high-order solitons within the current setup. High-order solitons (typically labeled by a soliton number $N$) are a cornerstone of nonlinear optics, characterized by their unique dynamics during propagation \cite{agrawal2000nonlinear}.
When a high-energy soliton pulse, \( A(t) = N \, \text{sech}(t) \), where $N$ is an integer soliton number, is seeded into a nonlinear fiber, spectro-temporal breather phenomena arise due to the imbalance between dispersion and self-phase modulation (SPM) during propagation along \( z \). The soliton period $z_0$ dictates the length for the pulse recovering to the initial state, which is evaluated by:
\begin{equation}
  z_0 = \frac{\pi}{2} L_d,
\end{equation}
where \( L_d = T_0^2 / |\beta_2| \) is the dispersion length. Specifically, the spectral evolution of a third-order soliton ($N$ = 3) over one soliton period is illustrated in Fig. \ref{F4}(b). In the time domain, the soliton contracts to a fraction of its initial width, splits into two pulses at \( z_0/2 \), and then recombines to its original form at \( z = z_0 \). This breathing pattern recurs over successive propagation segments \( z \).
Practically, capturing the breather pattern would traditionally require physically cutting the fiber or altering the pulse energy \cite{mollenauer1980experimental}, an irreversible and labor-intensive process. The P-CNN framework offers a more efficient method to balance dispersion and nonlinear effects, thus enabling an effective continuous scan of the output of the entire waveform over the position parameter \( z \).
In the following, we present the results observed from the P-CNN and also explain them using the theoretical model of high-order solitons.

We first set the target spectral pattern to waveform of third-order soliton and used the trained Net-S network alongside a strategic search approach to identify the optimal solution. The measured spectra, shown in the bottom panel of Fig.~\ref{F4}(b), exhibit a high degree of similarity to the target. Consistent with the simulations, the spectral intensity maintains its original profile initially, broadens around \(z_0/5\), and reconverges near \(z_0/4\). At the midpoint, \(z_0/2\), the spectrum splits, displaying conjugate dynamics for the latter  half of the propagation from \(z_0/2\) to \(z_0\).

To visualize their dynamics in the temporal domain, we used the SHG-FROG system, with results presented in Fig. \ref{F4}(e). The initial soliton has a pulse duration of approximately 850 fs, which narrows to a minimum of 23 fs at \(15/60z_0\), corresponding to a nonlinear compression factor of about 36. As the propagation continues, the pulse broadens and splits at \(z_0/2\), closely aligning with simulation results.

The same approach was applied to a fourth-order soliton, with results displayed in Fig.~\ref{F4}(c) and (f), respectively. In this case, the initial pulse duration, approximately 720 fs, reduces to a minimum of 22 fs, corresponding to a nonlinear compression factor of around 33. The temporal evolution of the fourth-order soliton is more complex, featuring multiple points of narrow pulse duration, consistent with its intricate spectro-temporal structure. (\textbf{Refer to the support flash for additional FROG results on high-order solitons}).

How do we prove that the dynamics simulated by P-CNN is of the solitonic kind? From the spectral phase and amplitude of the initial pulse optimized by the network and real system, we extract the temporal intensity, pulse duration (\(T_0\)) and peak power (\(P_k\)). Using these, we check the dispersion length (\(L_d=T_0^2/|\beta_2|\)) and nonlinear length (\(L_n = 1/(\gamma P_k)\)), which define the soliton number:

\begin{equation}\label{E-soliton}
  N^2 = \frac{L_d}{L_n} \approx T_0^2 P_k.
\end{equation}

Eq. \ref{E-soliton} emphasizes that the physical system must have the ability to control both the pulse duration \(T_0\) and the peak power \(P_k\), as these parameters directly influence the dispersion and nonlinear lengths. Therefore, maintenance of a consistent soliton number for each sequence is a key to avoiding destabilizing soliton dynamics \cite{agrawal2000nonlinear,malomed2006soliton}. Hence, to observe the full pulse evolution as a function of propagation length, P-CNN implements an effective length scan by varying the dispersion length \(L_d\), which can be self-consistently achieved by changing \(T_0\), \(P_k\), and the nonlinear length \(L_n\). 
Fig. \ref{F4}(d) presents the calculated \(L_d\) (red dots) and \(L_n\) (green dots) for $N$=4, based on the reconstructions in the pulse duration and peak power, demonstrating a periodic tendency. Notably, the reconstructed soliton number (green bars) shown in Fig. \ref{F4}(d) remains nearly constant, around 4, throughout the process, underscoring the importance of soliton number conservation for stable dynamics. \hyperref[sec:sup3]{\textbf{Further theoretical analysis on high-order solitons is provided in Supplementary-S6}}.
\section*{Discussion }

In this work, we introduced a \Red{physics-embedded} deep learning framework based on a convolutional neural network (CNN), termed P-CNN, designed for versatile spectro-temporal engineering of photonic wavepackets.
\Red{Unlike physics-informed neural networks (PINNs), which incorporate physical constraints into the loss function and are primarily designed to solve partial differential equations (PDEs) in the full space~\cite{karniadakis2021physics,lopez2023self}, our approach embeds physical feature directly into the input layer through a spectro-temporal correlation function—specifically the Wigner distribution, though other representations may also be applicable. This physics-embedded input structure boosts both CNN and FNN models to learn physically grounded features that capture the underlying ultrafast dynamics, leading to significantly lower prediction errors and improved robustness to noise compared to purely data-driven models.}

The framework was implemented and validated experimentally using an active pulse shaper combined with a highly nonlinear broadening section. The characterization of the output pulse via SHG-FROG facilitated wavepacket-based training data for the P-CNN. By analyzing the generated supercontinuum using two distinct networks, Net-D and Net-S, we successfully modeled the dynamics of dispersive waves (DW) and soliton waves (SW), respectively.

For Net-D, we achieved precise control over the DW properties, including center frequency, bandwidth, and pulse duration. Remarkably, a minimum infrared pulse duration of 12 fs—corresponding to sub-three optical cycles and near-transform-limited performance was achieved without external pulse compressors, enabled by the P-CNN’s ability to balance nonlinear broadening and pulse compression. \Blue{Pulses with even shorter durations could potentially be optimized by incorporating additional parameters into the training process, such as pump powers and dispersion losses.} For Net-S, we demonstrated on-demand generation of high-order solitons ($N = 3, 4$), capturing characteristic breather dynamics in the spectro-temporal domain, as predicted by the nonlinear Schrödinger equation.

These results highlight the exceptional capabilities of the P-CNN framework in enabling precise pulse manipulation across spectral and temporal domains. Traditionally, pulse shaping works at very limited spectral bandwidth. The P-CNN boosted nonlinear pulse shaper extends the broad spectrum band, opening new possibilities for targeted spectro-temporal engineering. \Blue{The pulse energy emitted from the current configuration is on the order of a few nanojoules. To extend the applicability of this approach to higher-energy regimes (e.g., $\mu$J-mJ level), integration with Hollow-Core Fibers (HCFs) or other large-mode-area nonlinear structures would be required—while still preserving the spectro-temporal control enabled by the proposed P-CNN-driven shaping method.}

An exciting avenue for future exploration is to combine spectro-temporal $(\omega,t)$ engineering with spatial degrees of freedom $(k,x)$ to create fully customized space-time designer wavepackets \cite{Jolly2020JO,Korman2022OL,chong2020generation}. This approach can be generalized to manipulate wavepackets in multidimensional correlation spaces, such as $(\omega, k)$ or $(t, x)$, offering transformative potential for nonlinear interactions and advanced beam shaping.

Beyond the demonstrations of GNLSE, the P-CNN framework presents a promising machine learning tool for solving other PDEs, which are fundamental in describing natural physical laws \cite{brunton2024promising}. Additionally, this approach has potential applications in accelerating optical computations in emerging fields such as AI photonics \cite{genty2021machine}, analog computing \cite{bandyopadhyay2024single}, and optical neuromorphic systems \cite{mcmahon2023physics,fischer2023neuromorphic}. Finally, the integration of physics-based insights into network design enhances model stability and convergence, particularly for large-scale data training systems. This robustness may help mitigate issues such as "model collapse" in large-language models, paving the way for more reliable and scalable AI-driven physical simulations \cite{shumailov2024ai}.

\section*{Experimental and Technical Design}
\phantomsection
\subsection*{Methods- 1: The setup and theoretical simulations for supercontinuum and few-cycle pulse}

\label{sec:methods1}

\begin{figure}
  \centering
\includegraphics[width=17 cm]{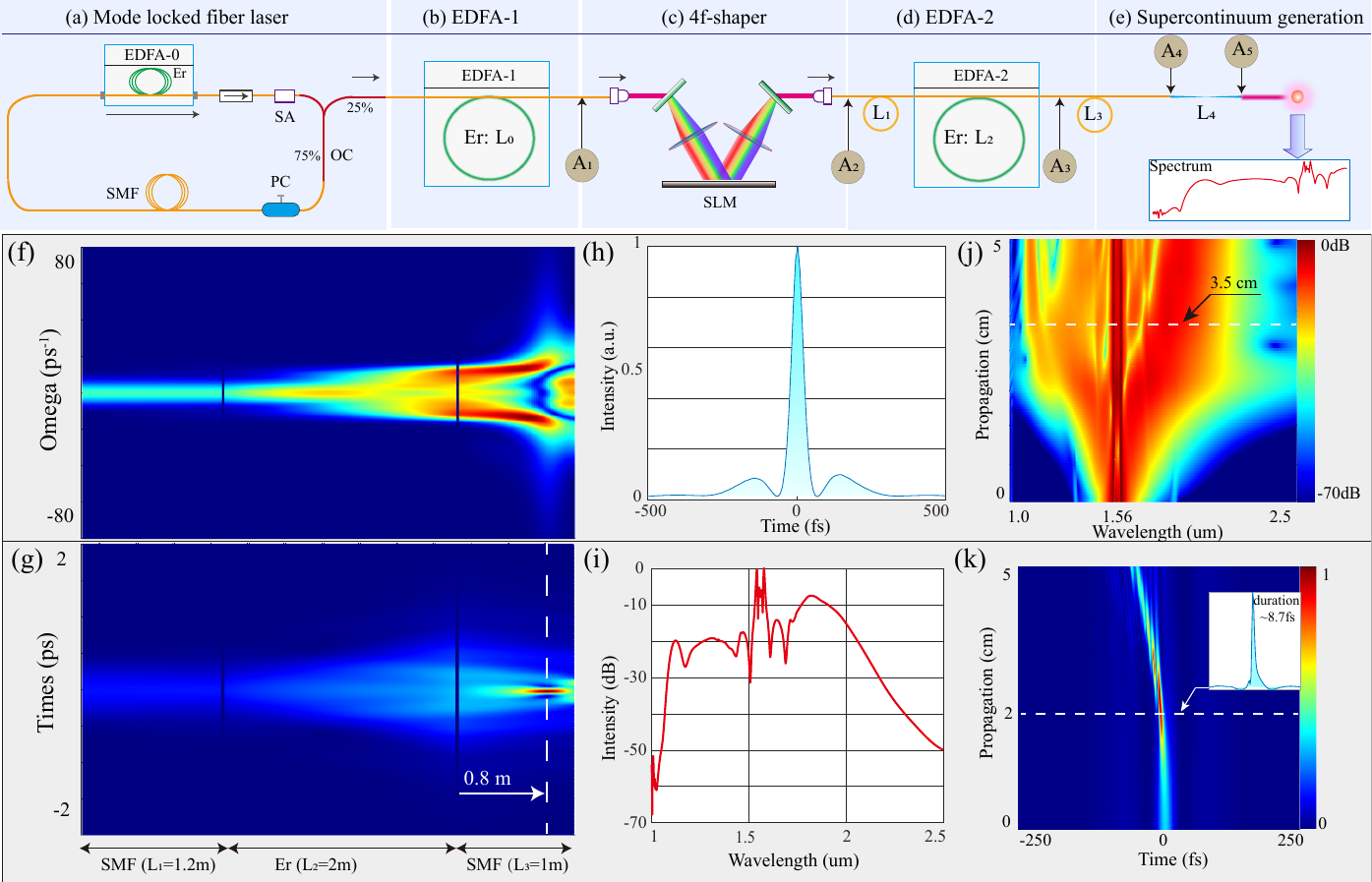}
  \caption{Setup and Simulations for Supercontinuum and Few-cycle pulse Generation. (a) Mode-locked fiber laser (MLF): The core laser source for generating initial soliton pulses. (b) EDFA-1: Includes an Erbium-doped fiber (EDF) \(L_0\) of 1.1 m, and single-mode fiber (SMF) $\sim$ 2.6 m.
(c) Pulse Shaper: Comprises two symmetric optical gratings (940 grooves /\ mm) and lenses (focal length $\sim$ 150 mm) positioned on either side of a Spatial Light Modulator (SLM, Santec SLM-100: 1440 $\times $1050 pixels).
(d) EDFA-2: Consists of a forward SMF \(L_1\) of 1.2 m, an Er-doped fiber \(L_2\) of 2 m, and an attached SMF \(L_3\) of 0.8 m.
(e) Supercontinuum generation: utilized with a Highly Nonlinear Fiber (HNF) of the length \(L_4\) of $\sim$ 4 cm.
(f) and (g) Dynamics of the Pulse and Spectral Intensity in simulations: Traces the evolution of the pulse and its spectral intensity through SMF (\(L_1\)), EDFA-2 (\(L_2\)), and SMF (\(L_3\)).
(h) Temporal Intensity Profile: Displays the temporal intensity at the position where SMF \(L_3\) is approximately 0.8 m, with the resultant pulse width around 46 fs.
(i) Calculated SC Spectrum: Shows the supercontinuum from 1 to 2.5 $\mathrm{\mu m}$ at a propagation length of 3.5 cm.
(j) and (k) Dynamics of SC: Depicts the spectral and temporal intensity variations of the SC along the HNF from 0 to 5 cm.}\label{Fig-setup}
\end{figure}

Fig. \ref{Fig-setup} illustrates the experimental setup for generating supercontinuum (SC) and few-cycle pulses. The setup comprises five main parts: (a) a homemade mode-locked fiber laser (MLF), (b) the first Erbium-Doped Fiber Amplifier (EDFA-1), (c) a 4f pulse shaper, (d) the second Erbium-Doped Fiber Amplifier (EDFA-2), and (e) a highly nonlinear fiber (HNF) used for SC emission.

The MLF is utilized to generate an infrared self-similar soliton pulse, essential for establishing self-similar amplification in both the temporal and spectral domains \cite{ilday2004self}. To achieve this, it is necessary to suppress the Kelly sidebands by optimizing the net dispersion within the cavity, adjusting the pump power in EDFA-0, and finely controlling the polarization by the polarization controller (PC) \cite{Liu2022SM}.

The soliton pulse from the MLF is measured using an efficient collinear Frequency-Resolved Optical Gating (FROG) system due to its weak pulse energy \cite{liu2021efficient}. The reconstructed pulse width is approximately 625 fs (\textbf{see Supplementary S5 for details}). Initially, this soliton pulse is stretched and amplified by the EDFA-1, as depicted in Fig \ref{Fig-setup}(b).  The amplified soliton pulse is directed through a 4f pulse shaper, as depicted in Fig. \ref{Fig-setup}(c). The pulse shaper assembly comprises two optical gratings, an infrared Spatial Light Modulator (SLM), and two Fourier lenses. After modulation by the 4f pulse shaper, the field can be mathematically represented by the following transfer equation:
 \begin{equation}\label{E-M2}
    A_2(t)=\mathcal{F}^{-1}[\mathcal{F}[A_1(t)] \cdot \exp{[i \tilde{\Phi}(\omega)]}]
 \end{equation}

Within the pulse shaper, one can intricately engineer the complex profile of \(\tilde{\Phi}(\omega)\), where the real part modulates the spectral phase, and the imaginary part manipulates the spectral amplitude \cite{walmsley2009characterization,bolduc2013exact,liu2019classical,Liu2022SM}. This precise control enables the fine-tuning of the pulse's spectral-temporal characteristics. The following EDFA-2, as depicted in Fig. \ref{Fig-setup} (d), is positioned at the output of the pulse shaper to enhance the pulse energy further. The dynamics of this amplification can be modeled by the
 following Nonlinear Schrödinger Equation:
 \begin{equation}\label{E-M1}
   i\frac{\partial}{\partial z} A = \frac{\beta_2}{2} \frac{\partial^2}{\partial t^2} A
   + i\frac{g(z)}{2} A - \gamma |A|^2 A
 \end{equation}
In this equation, the first term on the right side represents the Group Velocity Dispersion (GVD); the second term, \(g(z)\), accounts for the propagation-dependent gain profile; and the last term denotes the Kerr nonlinearity. This formula is also applicable to a single-mode fiber (SMF) by converting the gain term \(g(z)\) to a loss term, which we set to zero due to the short fiber length. The pulse energy is amplified, subsequently inducing a Self-Phase Modulation (SPM) effect because of the Kerr nonlinearity. The resulting pulse field envelope is denoted as \(A_3\) after traversing EDFA-2.

Since the output pulse from the EDFA-2 carries positive dispersion, an additional section of single-mode fiber (SMF), labeled \(L_3\) in Fig. \ref{Fig-setup} (d), is incorporated to provide compensating dispersion, effectively compressing the pulse.  The optimized pulse width at the position \(A_4\) is  $\sim$ 69.5 fs by FROG measurements.

Supercontinuum (SC) generation is achieved through the use of a HNF, with a selected length of $\sim$ 3.5 cm. The dynamics of the pulse in HNF can be described by the General Nonlinear Schrödinger Equation (GNLSE) \cite{dudley2006supercontinuum}, as outlined in Eq. 1 in the main text. Based on the GNLSE, we obtain the SC, denoted as \(A_5\) in Fig. \ref{Fig-setup}(e). This SC is subsequently analyzed using a spectrometer and a FROG system to measure its spectral and temporal profiles, respectively. To understand the pulse dynamics throughout the entire system, we perform numerical simulations using Eq. \ref{E-M1}-Eq. \ref{E-M2} along with the GNLSE, in which we use the reconstructed pulse profile from the FROG measurement on the soliton pulse as the input to the pulse shaper (\textbf{See Supp. S5 for FROG measurement details}).

Accounting for the contributions of the attached SMF and EDFA-2, the simulated spectral and temporal dynamics prior to the HNF, as depicted in Fig. \ref{Fig-setup}(f) and (g). Fig. \ref{Fig-setup}(h) presents the obtained pulse profiles with a width of approximately 46 fs for the position \(A_4\), which is marked by the dashed line in Fig. \ref{Fig-setup}(g).

When this pulse is input into the HNF, we observe the normalized distribution of the SC along the propagation length (0-5 cm) shown in Fig. \ref{Fig-setup}(j) on a logarithmic scale. Corresponding pulse intensity evaluations are displayed in Fig. \ref{Fig-setup}(k). As the propagation length increases, more complex nonlinear phenomena such as soliton fission and Raman-induced redshift may occur \cite{dudley2006supercontinuum}, causing the SC to broaden significantly from 1 to 2.5 $\mathrm{\mu m}$ after approximately 2 cm. In this position, the obtained minimum pulse duration is around 8.7 fs. For our setups, which focus primarily on dispersion wave components, the optimal position is set to 3.5 cm, indicated by the dashed line in  Fig. \ref{Fig-setup}(j). Corresponding one-dimensional spectrum is shown in Fig. \ref{Fig-setup}(i).

\phantomsection
\subsection*{Methods-2: The structure and performance of  \Red{Physics-embedded} Convolutional Neural Network (P-CNN)}
\label{sec:methods2}
\begin{figure}
  \centering
\includegraphics[width=17cm]{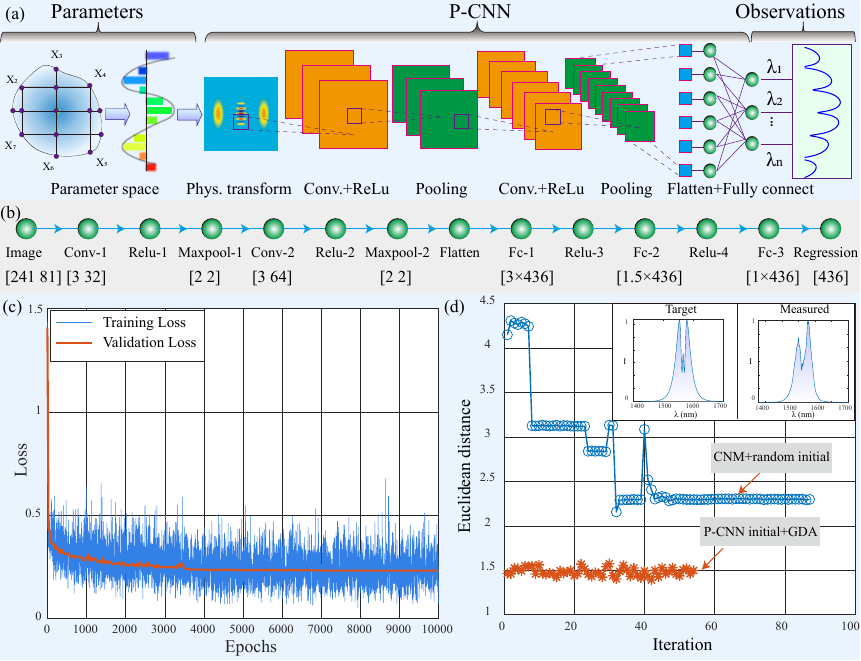}
  \caption{ Structure and Performance of P-CNN in Regression Tasks.
(a) Structure of the P-CNN: Illustrates the comprehensive architecture.
(b) Layer Actions: Details the specific operations performed at each layer of the CNN.
(c) Training Loss: Charts the loss throughout the training process, highlighting model convergence.
(d) Error Reduction in Euclidean distance : Compares the effectiveness of the constrained nonlinear multivariable (CNM) function and the gradient descent algorithm (GDA) in optimizing solutions, with initial conditions set to random values and inverted solutions derived from the trained CNN. The inset panel displays the target and measured spectra at the final iteration.
  }\label{Fig-CNN-1}
\end{figure}

Fig.~\ref{Fig-CNN-1}(a) illustrates the structure of the physics-involved convolutional neural network (CNN) utilized for the regression task in supercontinuum (SC) generation, with the detailed operations of each layer shown in Fig.~\ref{Fig-CNN-1}(b).
{
The parameter space $\{X_n\}$ defines the spectral phase ${\Phi}(\omega)$, which in turn determines the temporal electric field (after the pulse shaper) via Eq.~\ref{E-M2}. The input pulse (before the shaper) is reconstructed from FROG measurements, providing the initial spectral amplitude and phase.
By combining the initial field with the setting spectral phase obtained from $\{X_n\}$, we numerically compute the corresponding temporal field. The Wigner distribution is then calculated from this field using Eq.~\ref{E-3}.
This approach leverages the fact that the pulse shaper exhibits a reliable response between the programmed phase and the output field, as experimentally validated in our previous work~\cite{liu2023experimental}. This assumption allows for efficient dataset generation without requiring full experimental characterization for every instance.
The resulting Wigner function is cropped and downsampled from $8192 \times 8192$ to $241 \times 81$ pixels before being used as the input to the CNN.}

The first operation is a two-dimensional convolution layer, employing 32 filters of size $3\times3$, which are applied to the input image to extract features by emphasizing spatial hierarchies.
Following this, the Rectified Linear Unit (ReLU) layer acts as the second layer. It outputs the input directly if it is positive; otherwise, it produces zero. This activation function introduces non-linearity into the model, enabling the network to learn complex patterns and improve prediction accuracy. Subsequently, a `maxPooling' layer with a $2\times 2$ pool size is incorporated to reduce the spatial dimensions of the input feature maps. This reduction decreases both the computational complexity and the number of parameters within the network.

By maintaining a consistent layer structure to perform similar operations—convolution, ReLU activation, and max pooling, the network could capture more deep features from the input. Following this, a flatten layer is employed to transform multi-dimensional inputs into a one-dimensional vector. This transformation is essential for this application, as the output is a one-dimensional spectrum.

Subsequently, three `fullyConnectedLayer' s are utilized to gradually reduce the data dimensionality from $3N_{spectrum}$ to $N_{spectrum}$, where $N_{spectrum}$ represents the sampling number of the spectrum, noted as 436 for the Net-S. The architecture culminates with a `regressionLayer', which directly connects the target spectrum. This layer finalizes the predictive model by linking the learned features to the desired continuous outcome, facilitating accurate spectral regression based on the input characteristics. All these functions in CNN are assistant built-in Matlab deep learning toolbox.

For training this CNN, we utilized a dataset comprising 7,000 samples from our setup using the Latin Hypercube Sampling (LHS) method \cite{mckay2000comparison}.
This extensive dataset was processed using the `Adam' optimizer to train the network, in which the involved 75\% for training, 15\% for validation, and 10\% for testing.

 Fig. \ref{Fig-CNN-1}(c) illustrates the progression of training and validation loss throughout the training process. The validation loss stabilized after approximately 4,000 epochs, leveling off at around 0.23, indicating effective convergence of the network learning process. \textbf{See details for networks in supp. -S2.}

Fig. \ref{Fig-CNN-1}(d) illustrates the real-time iteration process within the optical setup for a specific target function, where the error is defined as the Elicude distance  between the normalized target and the measured spectrum. The blue dotted line represents the iterative process employing the constrained nonlinear multivariable (CNM) function to minimize the losses between the target and actual outputs \cite{byrd1999interior}. The distance stabilizes at 2.30 after approximately 50 iterations, starting from a random initial solution within the maximum range of \(X_n\).

In contrast, utilizing the gradient descent algorithm (GDA) results in a more rapid reduction of errors to 1.39, with three initial solutions provided by the trained CNN through inverse searching \textbf{(See supplementary- S3 for searching strategy)}. This approach demonstrates greater efficiency. Additionally, two inset panels in Fig. \ref{Fig-CNN-1}(d) display the target and measured spectra at the final iteration step, respectively, showing the effectiveness of the tuning process.

\section*{Data availability} The data that support the findings of this study are available from the corresponding authors, S.L. (dr.shilongliu@gmail.com) or D. V. S. (denis.seletskiy@polymtl.ca), upon reasonable request.



\section*{Acknowledgements}

\textbf{Appreciation:} The authors would like to thank Prof. Boris A. Malomed (Tel Aviv University)'s discussions in General Nonlinear Fractional Schrödinger Equation and high-order soliton. We would like to thank Prof. Ebrahim Karimi's SQO group for providing the infrared spatial light modulator. We appreciate the following colleagues in Polytechnique Montr\'{e}al for kind help during experiments and fruitful discussions:  Mr. Mika\'{e}l Leduc, Mr. Laurent Rivard, Mr. Marco Scaglia, Mr. \'{E}mile Dessureault, Mr. Gr\'{e}gory-Samuel Zagbayou, Mr. Rodrigo Itzamna Becerra Deana, Mr. Joseph Lamarre as well as Mr. \'{E}mile Jetzer. We thank Mrs. Christine Tao (Tempo. Optics Inc.) for improvements in figure aesthetics.

\textbf{Funding:}  This work was funded by Natural Sciences and Engineering Research Council of Canada (NSERC), via the Canada Research Chair program (CRC), and Fonds de Recherche du Qu\'{e}bec-Nature et Technologies (FRQNT), via Institut Transdisciplinaire d’Information Quantique (INTRIQ), and by the European Union's Horizon Europe Research and Innovation Programme under agreement 101070700 (project MIRAQLS). S.L. acknowledges the support PBEEE/Bourses de court s\'{e}jour de recherche ou perfectionnement, FRONT of Canada and Mitacs Accelerate Program.

\textbf{Competing interests:} Shilong Liu has financial interest in Tempo Optics Inc., which is developing photonic pulse shaper system.
The remaining authors declare no competing interests.

%
%
%
%

\begin{thebibliography}{10}
\urlstyle{rm}
\expandafter\ifx\csname url\endcsname\relax
  \def\url#1{\texttt{#1}}\fi
\expandafter\ifx\csname urlprefix\endcsname\relax\def\urlprefix{URL }\fi
\expandafter\ifx\csname doiprefix\endcsname\relax\def\doiprefix{DOI: }\fi
\providecommand{\bibinfo}[2]{#2}
\providecommand{\eprint}[2][]{\url{#2}}

\bibitem{Ranka2000OL}
\bibinfo{author}{Ranka, J.~K.}, \bibinfo{author}{Windeler, R.~S.} \&
  \bibinfo{author}{Stentz, A.~J.}
\newblock \bibinfo{journal}{\bibinfo{title}{Visible continuum generation in
  air--silica microstructure optical fibers with anomalous dispersion at 800
  nm}}.
\newblock {\emph{\JournalTitle{Opt. Lett.}}} \textbf{\bibinfo{volume}{25}},
  \bibinfo{pages}{25--27}, \doiprefix\url{10.1364/OL.25.000025}
  (\bibinfo{year}{2000}).

\bibitem{agrawal2000nonlinear}
\bibinfo{author}{Agrawal, G.~P.}
\newblock \bibinfo{title}{Nonlinear fiber optics}.
\newblock In \emph{\bibinfo{booktitle}{Nonlinear Science at the Dawn of the
  21st Century}}, \bibinfo{pages}{195--211} (\bibinfo{publisher}{Springer},
  \bibinfo{year}{2000}).

\bibitem{alfano2006supercontinuum}
\bibinfo{author}{Alfano, R.~R.}
\newblock \emph{\bibinfo{title}{The supercontinuum laser source: fundamentals
  with updated references}} (\bibinfo{publisher}{Springer},
  \bibinfo{year}{2006}).

\bibitem{dudley2006supercontinuum}
\bibinfo{author}{Dudley, J.~M.}, \bibinfo{author}{Genty, G.} \&
  \bibinfo{author}{Coen, S.}
\newblock \bibinfo{journal}{\bibinfo{title}{Supercontinuum generation in
  photonic crystal fiber}}.
\newblock {\emph{\JournalTitle{Reviews of modern physics}}}
  \textbf{\bibinfo{volume}{78}}, \bibinfo{pages}{1135} (\bibinfo{year}{2006}).

\bibitem{Sell2009OE}
\bibinfo{author}{Sell, A.}, \bibinfo{author}{Krauss, G.},
  \bibinfo{author}{Scheu, R.}, \bibinfo{author}{Huber, R.} \&
  \bibinfo{author}{Leitenstorfer, A.}
\newblock \bibinfo{journal}{\bibinfo{title}{8-fs pulses from a compact er:fiber
  system: quantitative modeling and experimental implementation}}.
\newblock {\emph{\JournalTitle{Opt. Express}}} \textbf{\bibinfo{volume}{17}},
  \bibinfo{pages}{1070--1077}, \doiprefix\url{10.1364/OE.17.001070}
  (\bibinfo{year}{2009}).

\bibitem{borondics2018supercontinuum}
\bibinfo{author}{Borondics, F.} \emph{et~al.}
\newblock \bibinfo{journal}{\bibinfo{title}{Supercontinuum-based fourier
  transform infrared spectromicroscopy}}.
\newblock {\emph{\JournalTitle{Optica}}} \textbf{\bibinfo{volume}{5}},
  \bibinfo{pages}{378--381} (\bibinfo{year}{2018}).

\bibitem{kaminski2008supercontinuum}
\bibinfo{author}{Kaminski, C.}, \bibinfo{author}{Watt, R.},
  \bibinfo{author}{Elder, A.}, \bibinfo{author}{Frank, J.} \&
  \bibinfo{author}{Hult, J.}
\newblock \bibinfo{journal}{\bibinfo{title}{Supercontinuum radiation for
  applications in chemical sensing and microscopy}}.
\newblock {\emph{\JournalTitle{Applied Physics B}}}
  \textbf{\bibinfo{volume}{92}}, \bibinfo{pages}{367--378}
  (\bibinfo{year}{2008}).

\bibitem{gundougdu2023self}
\bibinfo{author}{G{\"u}ndo{\u{g}}du, S.}, \bibinfo{author}{Virally, S.},
  \bibinfo{author}{Scaglia, M.}, \bibinfo{author}{Seletskiy, D.~V.} \&
  \bibinfo{author}{Moskalenko, A.~S.}
\newblock \bibinfo{journal}{\bibinfo{title}{Self-referenced subcycle metrology
  of quantum fields}}.
\newblock {\emph{\JournalTitle{Laser \& Photonics Reviews}}}
  \textbf{\bibinfo{volume}{17}}, \bibinfo{pages}{2200706}
  (\bibinfo{year}{2023}).

\bibitem{tu2013coherent}
\bibinfo{author}{Tu, H.} \& \bibinfo{author}{Boppart, S.~A.}
\newblock \bibinfo{journal}{\bibinfo{title}{Coherent fiber supercontinuum for
  biophotonics}}.
\newblock {\emph{\JournalTitle{Laser \& photonics reviews}}}
  \textbf{\bibinfo{volume}{7}}, \bibinfo{pages}{628--645}
  (\bibinfo{year}{2013}).

\bibitem{ji2021millimeter}
\bibinfo{author}{Ji, X.} \emph{et~al.}
\newblock \bibinfo{journal}{\bibinfo{title}{Millimeter-scale chip--based
  supercontinuum generation for optical coherence tomography}}.
\newblock {\emph{\JournalTitle{Science Advances}}}
  \textbf{\bibinfo{volume}{7}}, \bibinfo{pages}{eabg8869}
  (\bibinfo{year}{2021}).

\bibitem{Jones2000Sci}
\bibinfo{author}{Jones, D.~J.} \emph{et~al.}
\newblock \bibinfo{journal}{\bibinfo{title}{Carrier-envelope phase control of
  femtosecond mode-locked lasers and direct optical frequency synthesis}}.
\newblock {\emph{\JournalTitle{Science}}} \textbf{\bibinfo{volume}{288}},
  \bibinfo{pages}{635--639}, \doiprefix\url{10.1126/science.288.5466.635}
  (\bibinfo{year}{2000}).
\newblock
  \eprint{https://www.science.org/doi/pdf/10.1126/science.288.5466.635}.

\bibitem{Sinclair2015comb}
\bibinfo{author}{Sinclair, L.~C.} \emph{et~al.}
\newblock \bibinfo{journal}{\bibinfo{title}{Invited article: A compact
  optically coherent fiber frequency comb}}.
\newblock {\emph{\JournalTitle{Rev. Sci. Insrum.}}}
  \textbf{\bibinfo{volume}{86}}, \bibinfo{pages}{081301}
  (\bibinfo{year}{2015}).

\bibitem{Fehrenbacher2015Optica}
\bibinfo{author}{Fehrenbacher, D.} \emph{et~al.}
\newblock \bibinfo{journal}{\bibinfo{title}{Free-running performance and full
  control of a passively phase-stable er:fiber frequency comb}}.
\newblock {\emph{\JournalTitle{Optica}}} \textbf{\bibinfo{volume}{2}},
  \bibinfo{pages}{917--923}, \doiprefix\url{10.1364/OPTICA.2.000917}
  (\bibinfo{year}{2015}).

\bibitem{Krauss2010NatPhot}
\bibinfo{author}{Krauss, G.} \emph{et~al.}
\newblock \bibinfo{journal}{\bibinfo{title}{Synthesis of a single cycle of
  light with compact erbium-doped fibre technology}}.
\newblock {\emph{\JournalTitle{Nature Photonics}}}
  \textbf{\bibinfo{volume}{4}}, \bibinfo{pages}{33--36} (\bibinfo{year}{2010}).

\bibitem{shumakova2016multi}
\bibinfo{author}{Shumakova, V.} \emph{et~al.}
\newblock \bibinfo{journal}{\bibinfo{title}{Multi-millijoule few-cycle
  mid-infrared pulses through nonlinear self-compression in bulk}}.
\newblock {\emph{\JournalTitle{Nature communications}}}
  \textbf{\bibinfo{volume}{7}}, \bibinfo{pages}{12877} (\bibinfo{year}{2016}).

\bibitem{steinleitner2022single}
\bibinfo{author}{Steinleitner, P.} \emph{et~al.}
\newblock \bibinfo{journal}{\bibinfo{title}{Single-cycle infrared waveform
  control}}.
\newblock {\emph{\JournalTitle{Nature Photonics}}}
  \textbf{\bibinfo{volume}{16}}, \bibinfo{pages}{512--518}
  (\bibinfo{year}{2022}).

\bibitem{Lewenstein1994PRA}
\bibinfo{author}{Lewenstein, M.}, \bibinfo{author}{Balcou, P.},
  \bibinfo{author}{Ivanov, M.~Y.}, \bibinfo{author}{L'Huillier, A.} \&
  \bibinfo{author}{Corkum, P.~B.}
\newblock \bibinfo{journal}{\bibinfo{title}{Theory of high-harmonic generation
  by low-frequency laser fields}}.
\newblock {\emph{\JournalTitle{Phys. Rev. A}}} \textbf{\bibinfo{volume}{49}},
  \bibinfo{pages}{2117--2132}, \doiprefix\url{10.1103/PhysRevA.49.2117}
  (\bibinfo{year}{1994}).

\bibitem{Corkum2007NatPhys}
\bibinfo{author}{Corkum, P.~B.} \& \bibinfo{author}{Krausz, F.}
\newblock \bibinfo{journal}{\bibinfo{title}{Attosecond science}}.
\newblock {\emph{\JournalTitle{Nat. Phys.}}} \textbf{\bibinfo{volume}{3}},
  \bibinfo{pages}{381--387} (\bibinfo{year}{2007}).

\bibitem{Krausz2024RMP}
\bibinfo{author}{Krausz, F.}
\newblock \bibinfo{journal}{\bibinfo{title}{Nobel lecture: Sub-atomic
  motions}}.
\newblock {\emph{\JournalTitle{Rev. Mod. Phys.}}}
  \textbf{\bibinfo{volume}{96}}, \bibinfo{pages}{030502},
  \doiprefix\url{10.1103/RevModPhys.96.030502} (\bibinfo{year}{2024}).

\bibitem{peng2019attosecond}
\bibinfo{author}{Peng, P.}, \bibinfo{author}{Marceau, C.} \&
  \bibinfo{author}{Villeneuve, D.~M.}
\newblock \bibinfo{journal}{\bibinfo{title}{Attosecond imaging of molecules
  using high harmonic spectroscopy}}.
\newblock {\emph{\JournalTitle{Nature Reviews Physics}}}
  \textbf{\bibinfo{volume}{1}}, \bibinfo{pages}{144--155}
  (\bibinfo{year}{2019}).

\bibitem{li2020attosecond}
\bibinfo{author}{Li, J.} \emph{et~al.}
\newblock \bibinfo{journal}{\bibinfo{title}{Attosecond science based on high
  harmonic generation from gases and solids}}.
\newblock {\emph{\JournalTitle{Nature Communications}}}
  \textbf{\bibinfo{volume}{11}}, \bibinfo{pages}{2748} (\bibinfo{year}{2020}).

\bibitem{bres2023supercontinuum}
\bibinfo{author}{Br{\`e}s, C.-S.} \emph{et~al.}
\newblock \bibinfo{journal}{\bibinfo{title}{Supercontinuum in integrated
  photonics: generation, applications, challenges, and perspectives}}.
\newblock {\emph{\JournalTitle{Nanophotonics}}} \textbf{\bibinfo{volume}{12}},
  \bibinfo{pages}{1199--1244} (\bibinfo{year}{2023}).

\bibitem{lozovoy2004multiphoton}
\bibinfo{author}{Lozovoy, V.~V.}, \bibinfo{author}{Pastirk, I.} \&
  \bibinfo{author}{Dantus, M.}
\newblock \bibinfo{journal}{\bibinfo{title}{Multiphoton intrapulse
  interference. iv. ultrashort laser pulse spectral phase characterization and
  compensation}}.
\newblock {\emph{\JournalTitle{Opt. Lett.}}} \textbf{\bibinfo{volume}{29}},
  \bibinfo{pages}{775--777} (\bibinfo{year}{2004}).

\bibitem{xu2006quantitative}
\bibinfo{author}{Xu, B.}, \bibinfo{author}{Gunn, J.~M.}, \bibinfo{author}{Cruz,
  J. M.~D.}, \bibinfo{author}{Lozovoy, V.~V.} \& \bibinfo{author}{Dantus, M.}
\newblock \bibinfo{journal}{\bibinfo{title}{Quantitative investigation of the
  multiphoton intrapulse interference phase scan method for simultaneous phase
  measurement and compensation of femtosecond laser pulses}}.
\newblock {\emph{\JournalTitle{Journal of the Optical Society of America B}}}
  \textbf{\bibinfo{volume}{23}}, \bibinfo{pages}{750--759}
  (\bibinfo{year}{2006}).

\bibitem{monmayrant2010newcomer}
\bibinfo{author}{Monmayrant, A.}, \bibinfo{author}{Weber, S.~J.} \&
  \bibinfo{author}{Chatel, B.}
\newblock \bibinfo{journal}{\bibinfo{title}{A newcomer’s guide to ultrashort
  pulse shaping and characterization}}.
\newblock {\emph{\JournalTitle{J. Phys. B}}} \textbf{\bibinfo{volume}{43}},
  \bibinfo{pages}{103001} (\bibinfo{year}{2010}).

\bibitem{Carlson2017OL}
\bibinfo{author}{Carlson, D.~R.} \emph{et~al.}
\newblock \bibinfo{journal}{\bibinfo{title}{Self-referenced frequency combs
  using high-efficiency silicon-nitride waveguides}}.
\newblock {\emph{\JournalTitle{Opt. Lett.}}} \textbf{\bibinfo{volume}{42}},
  \bibinfo{pages}{2314--2317} (\bibinfo{year}{2017}).

\bibitem{beetar2020multioctave}
\bibinfo{author}{Beetar, J.~E.} \emph{et~al.}
\newblock \bibinfo{journal}{\bibinfo{title}{Multioctave supercontinuum
  generation and frequency conversion based on rotational nonlinearity}}.
\newblock {\emph{\JournalTitle{Science Advances}}}
  \textbf{\bibinfo{volume}{6}}, \bibinfo{pages}{eabb5375}
  (\bibinfo{year}{2020}).

\bibitem{elu2021seven}
\bibinfo{author}{Elu, U.} \emph{et~al.}
\newblock \bibinfo{journal}{\bibinfo{title}{Seven-octave high-brightness and
  carrier-envelope-phase-stable light source}}.
\newblock {\emph{\JournalTitle{Nature Photonics}}}
  \textbf{\bibinfo{volume}{15}}, \bibinfo{pages}{277--280}
  (\bibinfo{year}{2021}).

\bibitem{piccoli2021intense}
\bibinfo{author}{Piccoli, R.} \emph{et~al.}
\newblock \bibinfo{journal}{\bibinfo{title}{Intense few-cycle visible pulses
  directly generated via nonlinear fibre mode mixing}}.
\newblock {\emph{\JournalTitle{Nature Photonics}}}
  \textbf{\bibinfo{volume}{15}}, \bibinfo{pages}{884--889}
  (\bibinfo{year}{2021}).

\bibitem{genty2007fiber}
\bibinfo{author}{Genty, G.}, \bibinfo{author}{Coen, S.} \&
  \bibinfo{author}{Dudley, J.~M.}
\newblock \bibinfo{journal}{\bibinfo{title}{Fiber supercontinuum sources}}.
\newblock {\emph{\JournalTitle{JOSA B}}} \textbf{\bibinfo{volume}{24}},
  \bibinfo{pages}{1771--1785} (\bibinfo{year}{2007}).

\bibitem{genier2021ultra}
\bibinfo{author}{Genier, E.} \emph{et~al.}
\newblock \bibinfo{journal}{\bibinfo{title}{Ultra-flat, low-noise, and linearly
  polarized fiber supercontinuum source covering 670--1390 nm}}.
\newblock {\emph{\JournalTitle{Optics Letters}}} \textbf{\bibinfo{volume}{46}},
  \bibinfo{pages}{1820--1823} (\bibinfo{year}{2021}).

\bibitem{wetzel2018customizing}
\bibinfo{author}{Wetzel, B.} \emph{et~al.}
\newblock \bibinfo{journal}{\bibinfo{title}{Customizing supercontinuum
  generation via on-chip adaptive temporal pulse-splitting}}.
\newblock {\emph{\JournalTitle{Nature communications}}}
  \textbf{\bibinfo{volume}{9}}, \bibinfo{pages}{4884} (\bibinfo{year}{2018}).

\bibitem{sylvestre2021recent}
\bibinfo{author}{Sylvestre, T.} \emph{et~al.}
\newblock \bibinfo{journal}{\bibinfo{title}{Recent advances in supercontinuum
  generation in specialty optical fibers}}.
\newblock {\emph{\JournalTitle{JOSA B}}} \textbf{\bibinfo{volume}{38}},
  \bibinfo{pages}{F90--F103} (\bibinfo{year}{2021}).

\bibitem{tzang2018adaptive}
\bibinfo{author}{Tzang, O.}, \bibinfo{author}{Caravaca-Aguirre, A.~M.},
  \bibinfo{author}{Wagner, K.} \& \bibinfo{author}{Piestun, R.}
\newblock \bibinfo{journal}{\bibinfo{title}{Adaptive wavefront shaping for
  controlling nonlinear multimode interactions in optical fibres}}.
\newblock {\emph{\JournalTitle{Nature Photonics}}}
  \textbf{\bibinfo{volume}{12}}, \bibinfo{pages}{368--374}
  (\bibinfo{year}{2018}).

\bibitem{hary2023tailored}
\bibinfo{author}{Hary, M.} \emph{et~al.}
\newblock \bibinfo{journal}{\bibinfo{title}{Tailored supercontinuum generation
  using genetic algorithm optimized fourier domain pulse shaping}}.
\newblock {\emph{\JournalTitle{Optics Letters}}} \textbf{\bibinfo{volume}{48}},
  \bibinfo{pages}{4512--4515} (\bibinfo{year}{2023}).

\bibitem{lapre2023genetic}
\bibinfo{author}{Lapre, C.} \emph{et~al.}
\newblock \bibinfo{journal}{\bibinfo{title}{Genetic algorithm optimization of
  broadband operation in a noise-like pulse fiber laser}}.
\newblock {\emph{\JournalTitle{Scientific Reports}}}
  \textbf{\bibinfo{volume}{13}}, \bibinfo{pages}{1865} (\bibinfo{year}{2023}).

\bibitem{martins2022design}
\bibinfo{author}{Martins, G.~R.}, \bibinfo{author}{Silva, L.~C.},
  \bibinfo{author}{Segatto, M.~E.}, \bibinfo{author}{Rocha, H.~R.} \&
  \bibinfo{author}{Castellani, C.~E.}
\newblock \bibinfo{journal}{\bibinfo{title}{Design and analysis of recurrent
  neural networks for ultrafast optical pulse nonlinear propagation}}.
\newblock {\emph{\JournalTitle{Optics Letters}}} \textbf{\bibinfo{volume}{47}},
  \bibinfo{pages}{5489--5492} (\bibinfo{year}{2022}).

\bibitem{hoang2022optimizing}
\bibinfo{author}{Hoang, V.~T.} \emph{et~al.}
\newblock \bibinfo{journal}{\bibinfo{title}{Optimizing supercontinuum
  spectro-temporal properties by leveraging machine learning towards
  multi-photon microscopy}}.
\newblock {\emph{\JournalTitle{Frontiers in Photonics}}}
  \textbf{\bibinfo{volume}{3}}, \bibinfo{pages}{940902} (\bibinfo{year}{2022}).

\bibitem{shih2023maximizing}
\bibinfo{author}{Shih, M.} \emph{et~al.}
\newblock \bibinfo{journal}{\bibinfo{title}{Maximizing supercontinuum
  bandwidths in gas-filled hollow-core fibers using artificial neural
  networks}}.
\newblock {\emph{\JournalTitle{Journal of Applied Physics}}}
  \textbf{\bibinfo{volume}{133}} (\bibinfo{year}{2023}).

\bibitem{farfan2018femtosecond}
\bibinfo{author}{Farfan, C.~A.}, \bibinfo{author}{Epstein, J.} \&
  \bibinfo{author}{Turner, D.~B.}
\newblock \bibinfo{journal}{\bibinfo{title}{Femtosecond pulse compression using
  a neural-network algorithm}}.
\newblock {\emph{\JournalTitle{Optics Letters}}} \textbf{\bibinfo{volume}{43}},
  \bibinfo{pages}{5166--5169} (\bibinfo{year}{2018}).

\bibitem{boscolo2020artificial}
\bibinfo{author}{Boscolo, S.} \& \bibinfo{author}{Finot, C.}
\newblock \bibinfo{journal}{\bibinfo{title}{Artificial neural networks for
  nonlinear pulse shaping in optical fibers}}.
\newblock {\emph{\JournalTitle{Optics \& Laser Technology}}}
  \textbf{\bibinfo{volume}{131}}, \bibinfo{pages}{106439}
  (\bibinfo{year}{2020}).

\bibitem{salmela2022feed}
\bibinfo{author}{Salmela, L.} \emph{et~al.}
\newblock \bibinfo{journal}{\bibinfo{title}{Feed-forward neural network as
  nonlinear dynamics integrator for supercontinuum generation}}.
\newblock {\emph{\JournalTitle{Optics Letters}}} \textbf{\bibinfo{volume}{47}},
  \bibinfo{pages}{802--805} (\bibinfo{year}{2022}).

\bibitem{narhi2018machine}
\bibinfo{author}{N{\"a}rhi, M.} \emph{et~al.}
\newblock \bibinfo{journal}{\bibinfo{title}{Machine learning analysis of
  extreme events in optical fibre modulation instability}}.
\newblock {\emph{\JournalTitle{Nature communications}}}
  \textbf{\bibinfo{volume}{9}}, \bibinfo{pages}{4923} (\bibinfo{year}{2018}).

\bibitem{salmela2020machine}
\bibinfo{author}{Salmela, L.}, \bibinfo{author}{Lapre, C.},
  \bibinfo{author}{Dudley, J.~M.} \& \bibinfo{author}{Genty, G.}
\newblock \bibinfo{journal}{\bibinfo{title}{Machine learning analysis of rogue
  solitons in supercontinuum generation}}.
\newblock {\emph{\JournalTitle{Scientific Reports}}}
  \textbf{\bibinfo{volume}{10}}, \bibinfo{pages}{9596} (\bibinfo{year}{2020}).

\bibitem{genty2021machine}
\bibinfo{author}{Genty, G.} \emph{et~al.}
\newblock \bibinfo{journal}{\bibinfo{title}{Machine learning and applications
  in ultrafast photonics}}.
\newblock {\emph{\JournalTitle{Nat. Photonics}}} \textbf{\bibinfo{volume}{15}},
  \bibinfo{pages}{91--101} (\bibinfo{year}{2021}).

\bibitem{zuo2022deep}
\bibinfo{author}{Zuo, C.} \emph{et~al.}
\newblock \bibinfo{journal}{\bibinfo{title}{Deep learning in optical metrology:
  a review}}.
\newblock {\emph{\JournalTitle{Light: Science \& Applications}}}
  \textbf{\bibinfo{volume}{11}}, \bibinfo{pages}{39} (\bibinfo{year}{2022}).

\bibitem{walmsley2009characterization}
\bibinfo{author}{Walmsley, I.~A.} \& \bibinfo{author}{Dorrer, C.}
\newblock \bibinfo{journal}{\bibinfo{title}{Characterization of ultrashort
  electromagnetic pulses}}.
\newblock {\emph{\JournalTitle{Adv. Opt. Photonics}}}
  \textbf{\bibinfo{volume}{1}}, \bibinfo{pages}{308--437}
  (\bibinfo{year}{2009}).

\bibitem{najafabadi2024intensity}
\bibinfo{author}{Najafabadi, M.~S.} \emph{et~al.}
\newblock \bibinfo{journal}{\bibinfo{title}{Intensity correlations in the
  wigner representation}}.
\newblock {\emph{\JournalTitle{arXiv preprint arXiv:2407.12901}}}
  (\bibinfo{year}{2024}).

\bibitem{mckay2000comparison}
\bibinfo{author}{McKay, M.~D.}, \bibinfo{author}{Beckman, R.~J.} \&
  \bibinfo{author}{Conover, W.~J.}
\newblock \bibinfo{journal}{\bibinfo{title}{A comparison of three methods for
  selecting values of input variables in the analysis of output from a computer
  code}}.
\newblock {\emph{\JournalTitle{Technometrics}}} \textbf{\bibinfo{volume}{42}},
  \bibinfo{pages}{55--61} (\bibinfo{year}{2000}).

\bibitem{malomed2021optical}
\bibinfo{author}{Malomed, B.~A.}
\newblock \bibinfo{title}{Optical solitons and vortices in fractional media: A
  mini-review of recent results}.
\newblock In \emph{\bibinfo{booktitle}{Photonics}}, vol.~\bibinfo{volume}{8},
  \bibinfo{pages}{353} (\bibinfo{organization}{Multidisciplinary Digital
  Publishing Institute}, \bibinfo{year}{2021}).

\bibitem{liu2023experimental}
\bibinfo{author}{Liu, S.}, \bibinfo{author}{Zhang, Y.},
  \bibinfo{author}{Malomed, B.~A.} \& \bibinfo{author}{Karimi, E.}
\newblock \bibinfo{journal}{\bibinfo{title}{Experimental realisations of the
  fractional schr{\"o}dinger equation in the temporal domain}}.
\newblock {\emph{\JournalTitle{Nature Communications}}}
  \textbf{\bibinfo{volume}{14}}, \bibinfo{pages}{222} (\bibinfo{year}{2023}).

\bibitem{kottig2017mid}
\bibinfo{author}{K{\"o}ttig, F.} \emph{et~al.}
\newblock \bibinfo{journal}{\bibinfo{title}{Mid-infrared dispersive wave
  generation in gas-filled photonic crystal fibre by transient
  ionization-driven changes in dispersion}}.
\newblock {\emph{\JournalTitle{Nature communications}}}
  \textbf{\bibinfo{volume}{8}}, \bibinfo{pages}{813} (\bibinfo{year}{2017}).

\bibitem{moussa2023observation}
\bibinfo{author}{Moussa, H.} \emph{et~al.}
\newblock \bibinfo{journal}{\bibinfo{title}{Observation of temporal reflection
  and broadband frequency translation at photonic time interfaces}}.
\newblock {\emph{\JournalTitle{Nature Physics}}} \textbf{\bibinfo{volume}{19}},
  \bibinfo{pages}{863--868} (\bibinfo{year}{2023}).

\bibitem{anderson1983nonlinear}
\bibinfo{author}{Anderson, D.} \& \bibinfo{author}{Lisak, M.}
\newblock \bibinfo{journal}{\bibinfo{title}{Nonlinear asymmetric self-phase
  modulation and self-steepening of pulses in long optical waveguides}}.
\newblock {\emph{\JournalTitle{Physical Review A}}}
  \textbf{\bibinfo{volume}{27}}, \bibinfo{pages}{1393} (\bibinfo{year}{1983}).

\bibitem{mollenauer1980experimental}
\bibinfo{author}{Mollenauer, L.~F.}, \bibinfo{author}{Stolen, R.~H.} \&
  \bibinfo{author}{Gordon, J.~P.}
\newblock \bibinfo{journal}{\bibinfo{title}{Experimental observation of
  picosecond pulse narrowing and solitons in optical fibers}}.
\newblock {\emph{\JournalTitle{Phys. Rev. Lett.}}}
  \textbf{\bibinfo{volume}{45}}, \bibinfo{pages}{1095} (\bibinfo{year}{1980}).

\bibitem{malomed2006soliton}
\bibinfo{author}{Malomed, B.~A.}
\newblock \emph{\bibinfo{title}{Soliton management in periodic systems}}
  (\bibinfo{publisher}{Springer Science \& Business Media},
  \bibinfo{year}{2006}).

\bibitem{karniadakis2021physics}
\bibinfo{author}{Karniadakis, G.~E.} \emph{et~al.}
\newblock \bibinfo{journal}{\bibinfo{title}{Physics-informed machine
  learning}}.
\newblock {\emph{\JournalTitle{Nature Reviews Physics}}}
  \textbf{\bibinfo{volume}{3}}, \bibinfo{pages}{422--440}
  (\bibinfo{year}{2021}).

\bibitem{lopez2023self}
\bibinfo{author}{Lopez-Pastor, V.} \& \bibinfo{author}{Marquardt, F.}
\newblock \bibinfo{journal}{\bibinfo{title}{Self-learning machines based on
  hamiltonian echo backpropagation}}.
\newblock {\emph{\JournalTitle{Physical Review X}}}
  \textbf{\bibinfo{volume}{13}}, \bibinfo{pages}{031020}
  (\bibinfo{year}{2023}).

\bibitem{Jolly2020JO}
\bibinfo{author}{Jolly, S.~W.}, \bibinfo{author}{Gobert, O.} \&
  \bibinfo{author}{Quéré, F.}
\newblock \bibinfo{journal}{\bibinfo{title}{Spatio-temporal characterization of
  ultrashort laser beams: a tutorial}}.
\newblock {\emph{\JournalTitle{Journal of Optics}}}
  \textbf{\bibinfo{volume}{22}}, \bibinfo{pages}{103501},
  \doiprefix\url{10.1088/2040-8986/abad08} (\bibinfo{year}{2020}).

\bibitem{Korman2022OL}
\bibinfo{author}{Korman, S.}, \bibinfo{author}{Bahar, E.},
  \bibinfo{author}{Arieli, U.} \& \bibinfo{author}{Suchowski, H.}
\newblock \bibinfo{journal}{\bibinfo{title}{Spatio-temporal ultrafast pulse
  shaping at the femtosecond-nanometer scale}}.
\newblock {\emph{\JournalTitle{Opt. Lett.}}} \textbf{\bibinfo{volume}{47}},
  \bibinfo{pages}{4279--4282}, \doiprefix\url{10.1364/OL.461953}
  (\bibinfo{year}{2022}).

\bibitem{chong2020generation}
\bibinfo{author}{Chong, A.}, \bibinfo{author}{Wan, C.}, \bibinfo{author}{Chen,
  J.} \& \bibinfo{author}{Zhan, Q.}
\newblock \bibinfo{journal}{\bibinfo{title}{Generation of spatiotemporal
  optical vortices with controllable transverse orbital angular momentum}}.
\newblock {\emph{\JournalTitle{Nat. Photonics}}} \textbf{\bibinfo{volume}{14}},
  \bibinfo{pages}{350--354} (\bibinfo{year}{2020}).

\bibitem{brunton2024promising}
\bibinfo{author}{Brunton, S.~L.} \& \bibinfo{author}{Kutz, J.~N.}
\newblock \bibinfo{journal}{\bibinfo{title}{Promising directions of machine
  learning for partial differential equations}}.
\newblock {\emph{\JournalTitle{Nature Computational Science}}}
  \bibinfo{pages}{1--12} (\bibinfo{year}{2024}).

\bibitem{bandyopadhyay2024single}
\bibinfo{author}{Bandyopadhyay, S.} \emph{et~al.}
\newblock \bibinfo{journal}{\bibinfo{title}{Single-chip photonic deep neural
  network with forward-only training}}.
\newblock {\emph{\JournalTitle{Nature Photonics}}} \bibinfo{pages}{1--9}
  (\bibinfo{year}{2024}).

\bibitem{mcmahon2023physics}
\bibinfo{author}{McMahon, P.~L.}
\newblock \bibinfo{journal}{\bibinfo{title}{The physics of optical computing}}.
\newblock {\emph{\JournalTitle{Nature Reviews Physics}}}
  \textbf{\bibinfo{volume}{5}}, \bibinfo{pages}{717--734}
  (\bibinfo{year}{2023}).

\bibitem{fischer2023neuromorphic}
\bibinfo{author}{Fischer, B.} \emph{et~al.}
\newblock \bibinfo{journal}{\bibinfo{title}{Neuromorphic computing via
  fission-based broadband frequency generation}}.
\newblock {\emph{\JournalTitle{Advanced Science}}}
  \textbf{\bibinfo{volume}{10}}, \bibinfo{pages}{2303835}
  (\bibinfo{year}{2023}).

\bibitem{shumailov2024ai}
\bibinfo{author}{Shumailov, I.} \emph{et~al.}
\newblock \bibinfo{journal}{\bibinfo{title}{Ai models collapse when trained on
  recursively generated data}}.
\newblock {\emph{\JournalTitle{Nature}}} \textbf{\bibinfo{volume}{631}},
  \bibinfo{pages}{755--759} (\bibinfo{year}{2024}).

\bibitem{ilday2004self}
\bibinfo{author}{Ilday, F.}, \bibinfo{author}{Buckley, J.},
  \bibinfo{author}{Clark, W.} \& \bibinfo{author}{Wise, F.}
\newblock \bibinfo{journal}{\bibinfo{title}{Self-similar evolution of parabolic
  pulses in a laser}}.
\newblock {\emph{\JournalTitle{Phys. Rev. Lett.}}}
  \textbf{\bibinfo{volume}{92}}, \bibinfo{pages}{213902}
  (\bibinfo{year}{2004}).

\bibitem{Liu2022SM}
\bibinfo{author}{Liu, S.}, \bibinfo{author}{Cui, Y.}, \bibinfo{author}{Karimi,
  E.} \& \bibinfo{author}{Malomed, B.~A.}
\newblock \bibinfo{journal}{\bibinfo{title}{On-demand harnessing of photonic
  soliton molecules}}.
\newblock {\emph{\JournalTitle{Optica}}} \textbf{\bibinfo{volume}{9}},
  \bibinfo{pages}{240--250} (\bibinfo{year}{2022}).

\bibitem{liu2021efficient}
\bibinfo{author}{Liu, S.}, \bibinfo{author}{Cui, Y.}, \bibinfo{author}{Zhou,
  Z.} \& \bibinfo{author}{Karimi, E.}
\newblock \bibinfo{title}{An efficient collinear frog system to character the
  ultrafast infrared laser pulse}.
\newblock In \emph{\bibinfo{booktitle}{2021 Photonics \& Electromagnetics
  Research Symposium (PIERS)}}, \bibinfo{pages}{2690--2694}
  (\bibinfo{organization}{IEEE}, \bibinfo{year}{2021}).

\bibitem{bolduc2013exact}
\bibinfo{author}{Bolduc, E.}, \bibinfo{author}{Bent, N.},
  \bibinfo{author}{Santamato, E.}, \bibinfo{author}{Karimi, E.} \&
  \bibinfo{author}{Boyd, R.~W.}
\newblock \bibinfo{journal}{\bibinfo{title}{Exact solution to simultaneous
  intensity and phase encryption with a single phase-only hologram}}.
\newblock {\emph{\JournalTitle{Opt. Lett.}}} \textbf{\bibinfo{volume}{38}},
  \bibinfo{pages}{3546--3549} (\bibinfo{year}{2013}).

\bibitem{liu2019classical}
\bibinfo{author}{Liu, S.-L.} \emph{et~al.}
\newblock \bibinfo{journal}{\bibinfo{title}{Classical analogy of a cat state
  using vortex light}}.
\newblock {\emph{\JournalTitle{Communications Physics}}}
  \textbf{\bibinfo{volume}{2}}, \bibinfo{pages}{1--9} (\bibinfo{year}{2019}).

\bibitem{byrd1999interior}
\bibinfo{author}{Byrd, R.~H.}, \bibinfo{author}{Hribar, M.~E.} \&
  \bibinfo{author}{Nocedal, J.}
\newblock \bibinfo{journal}{\bibinfo{title}{An interior point algorithm for
  large-scale nonlinear programming}}.
\newblock {\emph{\JournalTitle{SIAM Journal on Optimization}}}
  \textbf{\bibinfo{volume}{9}}, \bibinfo{pages}{877--900}
  (\bibinfo{year}{1999}).

\end{thebibliography}

\end{document}